\documentclass[5p,twocolumn,10p,times]{elsarticle}

\usepackage{amssymb}
\usepackage{url}
\usepackage{appendix}
\usepackage{amsmath}
\usepackage{hyperref}
\usepackage{graphicx}
\usepackage{subfigure}
\usepackage{verbatim}
\usepackage{ulem}
\usepackage[ruled]{algorithm2e} 
\usepackage{marvosym}
\usepackage{algorithmic}
\usepackage{multirow}
\usepackage{caption}
\usepackage{amsmath}
\usepackage{wrapfig}
\usepackage{ctable}
\usepackage{comment}
\biboptions{sort&compress}
\graphicspath{{images/}}

\newcommand{\rev}[2]{#2}

\journal{Composites Part B: Engineering}

\begin{document}
\begin{frontmatter}

\title{Toolpath Generation for High Density Spatial Fiber Printing Guided by Principal Stresses}

\author{Tianyu Zhang\textsuperscript{a,*}}
\author{Tao Liu\textsuperscript{a,*}}
\author{Neelotpal Dutta\textsuperscript{a}}
\author{Yongxue Chen\textsuperscript{a}}
\author{Renbo Su\textsuperscript{a}}
\author{Zhizhou Zhang\textsuperscript{a}}
\author{Weiming Wang\textsuperscript{a}}
\author{Charlie C.L. Wang\textsuperscript{a,**}}
\cortext[cor1]{Joint first authors}
\cortext[cor2]{Corresponding author. E-mail: changling.wang@manchester.ac.uk}
\affiliation{organization={Department of Mechanical and Aerospace Engineering},
            addressline={The University of Manchester},
            country={United Kingdom}}

\begin{abstract}
While multi-axis 3D printing can align continuous fibers along principal stresses in continuous fiber-reinforced thermoplastic (CFRTP) composites to enhance mechanical strength, existing methods have difficulty generating toolpaths with high fiber coverage. This is mainly due to the orientation consistency constraints imposed by vector-field-based methods and the turbulent stress fields around stress concentration regions. This paper addresses these challenges by introducing a 2-RoSy representation for computing the direction field, which is then converted into a periodic scalar field to generate partial iso-curves for fiber toolpaths with nearly equal hatching distance. To improve fiber coverage in stress-concentrated regions, such as around holes, we extend the quaternion-based method for curved slicing by incorporating winding compatibility considerations. Our proposed method can achieve toolpaths coverage between 87.5\% and 90.6\% by continuous fibers with 1.1mm width. Specimens fabricated using our toolpaths show up to 84.6\% improvement in failure load and 54.4\% increase in stiffness when compared to the results obtained from multi-axis 3D printing with sparser fibers.
\end{abstract}


\end{frontmatter}

\section{Introduction}\label{secIntroduction}
Additive manufacturing (AM) for continuous fiber-reinforced thermoplastic (CFRTP) composites offers unprecedented opportunities to rapidly develop next-generation high-performance composites with selective and spatially distributed reinforcement. Due to their high strength and lightweight characteristics, continuous fiber-reinforced polymers have shown great potential in a wide range of applications, such as aerospace structural parts~\cite{Sultana_CPA20} and wind turbine blades~\cite{Wang_ADDMA20}. The CFRTP-AM technology was developed based on Fused Deposition Modeling (FDM), which enables direct 3D manufacturing without the use of molds, using polymers such as polylactic acid (PLA)~\cite{Wang_ADDMA21, Li_CPB21}, polyamide (PA)~\cite{Wang_CS21} as the matrix material, and fibers such as carbon fiber~\cite{Goh_CPB21, Almeida_CS24}, glass fiber~\cite{Khosravani_TAFM22}, liquid crystal~\cite{Wang_NC23}, natural fiber~\cite{Awais_FP20} are usually employed as reinforcement materials. 

The \rev{density}{volume fraction} of fibers in composite materials is a critical metric in the industry, as it directly impacts the mechanical properties of fiber-reinforced composites~\cite{Ucsun_ADDMA24}. Significant attention has also been devoted to research on optimizing fiber-reinforcement effectiveness. Aligning fibers along the direction of maximum principal stress is widely recognized as a method to maximize reinforcement~\cite{Heitkamp_PAM23}. However, few methods are currently available that can both align fibers along the principal stress direction and simultaneously increase the \rev{density}{volume fraction} of fibers. \rev{}{In the rest of this paper, we borrow the term 'density' to refer specifically to the volume fraction of fiber.} 

\subsection{Related Work}
The focus of our paper, toolpath planning, plays a key role in connecting the sliced curved layers to the physical material placement. The existing toolpath generation approaches for CFRTP composites can be classified into two major groups.

\subsubsection{Geometric rule-based method} 
In previous research works, several types of toolpaths have been introduced for the fabrication of continuous fiber-reinforced thermoplastic (CFRTP) composites. The commonly used CFRTP printing patterns include zigzag paths~\cite{Wang_NC24, Zhu_PAT23}, sinusoidal-path~\cite{Shang_CST20}, spiral paths~\cite{Akhoundi_JEM20}, cellular paths~\cite{Cheng_CPA24, Cheng_CC23}, contour-parallel paths~\cite{Fernandes_AST21,Nguyen_JBSMSE23} and hybrid paths~\cite{Melenka_CS16}. In industry, Markforged's Eiger software~\cite{Eiger} uses a zigzag-contour pattern for the printing process. Geometry-based methods have advantages in their simplicity, adaptability, and ease of controlling fiber volume ratio, enabling high fiber density or coverage rates that enhance structural strength. However, these methods focus primarily on geometric features and often neglect the integration of loading information such as stresses in toolpath generation, which reduces the potential for strength improvements. Moreover, these approaches are largely confined to planar-layer-based printing techniques, further limiting their effectiveness in optimizing fiber reinforcement.

\subsubsection{Stress-aware fiber toolpath generation}
Fibers demonstrate exceptional strength and stiffness properties along their longitudinal direction. However, these properties significantly diminish when loads are applied vertically to the fiber axis. The previous research~\cite{Suzuki_CMAME91} has already proved that optimal fiber reinforcement can be achieved by aligning fibers along with the principal stress directions. The underlying concept of this approach is to minimize shear stresses, while maximizing normal stresses, thereby effectively leveraging the longitudinal strength properties of the fibers. Two approaches are typically considered for utilizing stress field information obtained from finite element analysis (FEA).

One widely employed strategy is to design the toolpaths by tracing the stress lines. Xia et al.~\cite{Xia_ADDMA20} implemented this strategy to generate the toolpaths by tracing the directional change of maximal stresses from element to element incrementally. Li et al.~\cite{Li_CPB20} decompose the topology optimization outcomes of structures primarily into two components: scaffold beams and branches. The beams are manufactured using anisotropic continuous fiber-reinforced filament aligned along the direction of the boundary.  A load-dependent paths planning (LPP) method is proposed in~\cite{Wang_CPA21} to generate printing toolpaths for CFRTP composites, which are traced along the load transmission path through the application of the Stress Vector Tracing (SVT) algorithm. Liu et al.~\cite{Liu_MD24} presented an enhanced tracing method that utilizes principal stress direction to control and smooth toolpaths while also maintaining adherence to the minimum allowable bead width constraint. Although the stress line extraction method is simple and easy to implement, it requires a lot of pre-processing operations, such as flipping the stress field vector to make the consistent orientation of the principal stresses. Moreover, in the turbulent areas of stress fields, stress line extraction is prone to failure, resulting in low robustness.

The other strategy is to implicitly extract the toolpaths from the stress fields or other optimized fields. Sugiyama et al.~\cite{Sugiyama_CST20} refined curved fiber trajectories through iterative adjustments of fiber orientation to align them with the principal stress direction obtained from stress field calculations. Anisoprint~\cite{Anisoprint} applied the stress-field guided toolpath to their products that have been applied to industrial applications. A continuous fiber-reinforced toolpath can be produced by extracting the iso-curves of a scalar field~\cite{Chen_ADDMA22}. The scalar field employed in their method has considered both the direction and magnitude of the maximal stresses in critical areas and the continuity of toolpaths near the boundaries. Fang et al.~\cite{Fang_ADDMA24} introduced a spatial printing technique for manufacturing CFRTP composites, with a focus on two key aspects: 1) aligning continuous fibers along with maximal stress directions in crucial areas, and 2) linking multiple load-bearing regions using continuous fibers. They utilized traced stress lines to optimize the scalar field, from which the continuous carbon fiber toolpaths were extracted. \rev{}{Ren et al.~\cite{Ren_CMAME24} and Liu et al.~\cite{Liu_ADDMA23} employed the `wave projection method' to link the vector field with the phase field (or scalar field), facilitating the toolpath generation. However, their methods using a scalar-field with wave functions cannot represent the discontinuity of toolpaths, which prevent the generation of uniform hatching space, which is essential for maintaining a high fiber coverage rate.} In summary, although field-based methods can handle the principal stresses following requirements in general while incorporating manufacturing constraints, they usually produce toolpaths with sparse density thereby limiting the mechanical strength of fabricated models.

\subsection{Challenges of Stress-Aware Fiber Placement}\label{subSec:ProblemState}

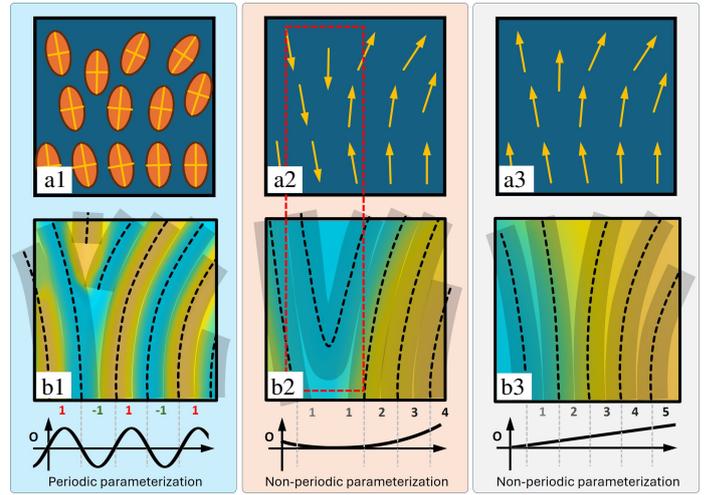
\begin{figure}[t]
\centering  
\includegraphics[width=\linewidth]{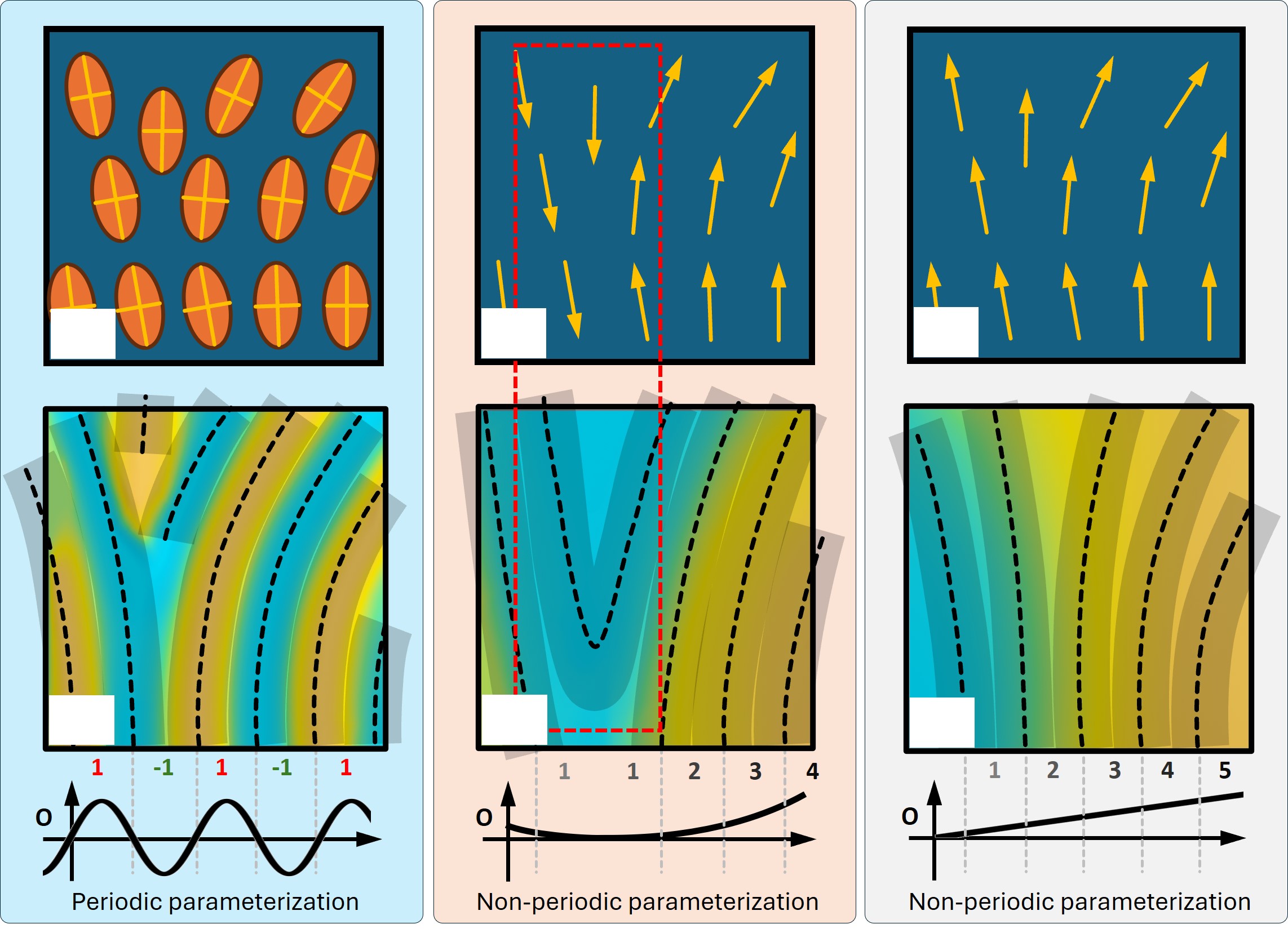}\vspace{-5pt}
\put(-247,117){\footnotesize \color{black}(a1)}
\put(-161,117){\footnotesize \color{black}(a2)}
\put(-75,117){\footnotesize \color{black}(a3)}
\put(-248,39){\footnotesize \color{black}(b1)}
\put(-162,39){\footnotesize \color{black}(b2)}
\put(-76,39){\footnotesize \color{black}(b3)}
\caption{Challenges associated with stress-aware fiber placement in prior methods. (a1) The stress tensors and (a2) the maximal principal stresses obtained by eigenvalue decomposition. The toolpath generated based on the field which exhibits directional ambiguity~\cite{Fang_SIGA20}, as depicted in (b2), deviates from the actual stress field, particularly in regions with significant directional changes -- see the region highlighted by red dash lines. (a3) Even when the direction of the stress field is corrected to obtain a vector field with uniform orientation, the generated path (b3) fails to fully cover the surface layer~\cite{Fang_ADDMA24} -- see those regions covered by the gray shadows. By contrast, our proposed method directly generates a periodic scalar field (b1), enabling the computation of fiber paths with near-constant hatching distance and achieving a dense, uniform fiber arrangement.
}\label{fig:problem_statement}
\vspace{-5pt}
\end{figure}

In stress-based toolpath planning, the stress tensors can be obtained from FEA as shown in Fig.~\ref{fig:problem_statement}(a1). The maximum principal stress $\sigma_{max}$ in each element can then be extracted using eigenvalue decomposition, as depicted in Fig.~\ref{fig:problem_statement}(a2). However, the resulting principal stress field (as a vector field) poses significant challenges for toolpath generation due to the following reasons.

\begin{figure}[t]
\centering
\includegraphics[width=0.85\linewidth]{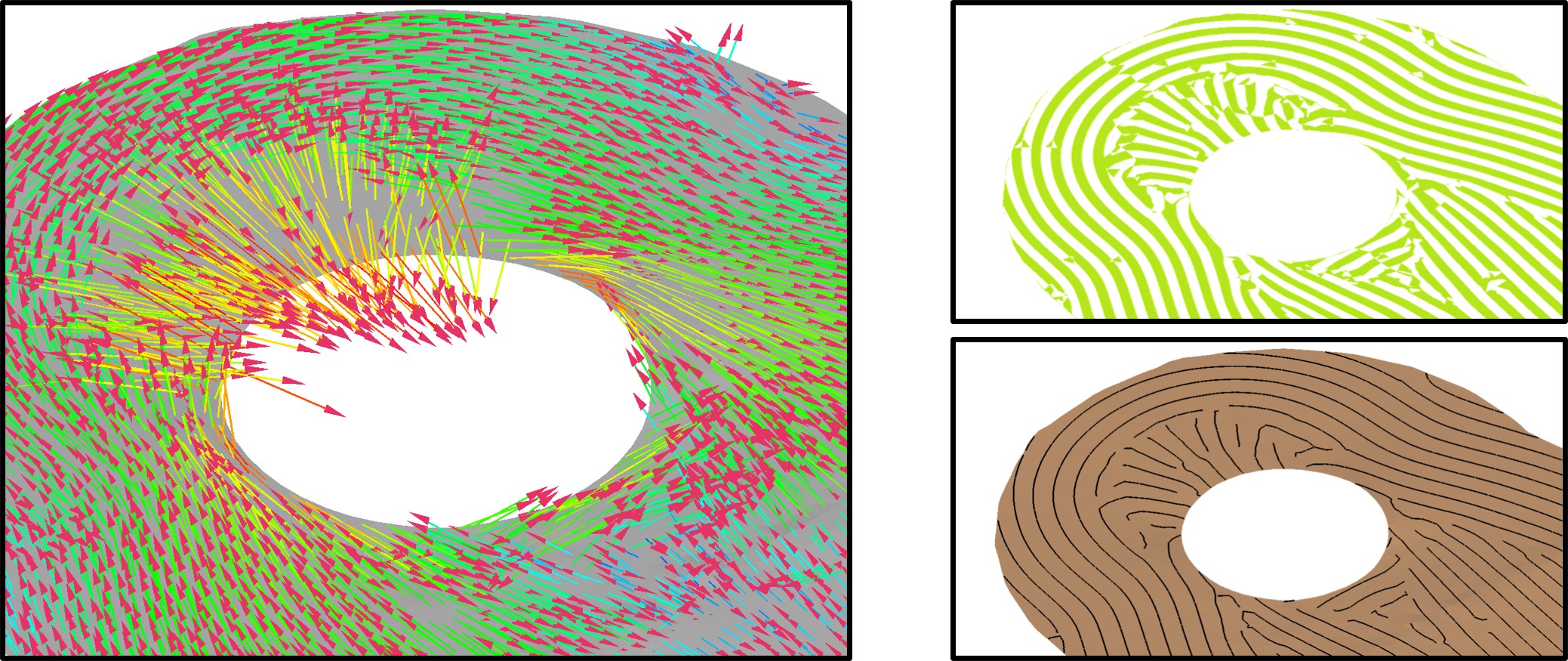}
\put(-231,0){\footnotesize \color{black}(a)}
\put(-98,48){\footnotesize \color{black}(b)}
\put(-98,0){\footnotesize \color{black}(c)}
\caption{The impact of turbulent region in the stress field on toolpath generation: (a) Stress concentration near the load-bearing holes induces turbulence in the stress field; (b \& c) Periodic parameterization results obtained using the original stress field as input, with the corresponding toolpaths extracted from the strip patterns.
}\label{fig:Turbulence}
\end{figure}

\begin{itemize} 
\item \textbf{Directional Ambiguity:} Vectors of maximal principal stresses in a stress field may have their orientations randomly flipped due to the arbitrary sign given in eigenvalue decomposition (Fig.~\ref{fig:problem_statement}(a2)). Simply applying a smoothing operator to resolve these ambiguities often leads to vanishing issues, where the vector length approaches zero. An alternative method of incrementally flipping vector orientations from a fixed anchor point fails to ensure global smoothness and coherence across the field (especially in turbulent areas). Similar challenges have also been reported in the surface reconstruction research where the consistent orientation of normal vectors is required ~\cite{Hoppe_SIG92,Liu_CG10}.

\item \textbf{Low Fiber Coverage:} 
Even after adjusting a stress field into a uniformly oriented vector field, as shown in Fig.~\ref{fig:problem_statement}(a3), the toolpaths generated by converting the vector field into a scalar field and then extracting the iso-curves fail to achieve dense surface coverage. This issue arises from the non-equidistant spacing of iso-curves extracted from the general scalar field. Due to the required minimal distance between neighboring fibers\footnote{Continuous fibers employed in CFRTP composites usually have a fixed width and are not allowed to overlap during deposition.}, existing stress-aware toolpath generation methods can only achieve dense fiber placement in narrow regions or areas of stress concentration, as seen in Fig.~\ref{fig:problem_statement}(b3). As a result, a large part of the model remains uncovered by the fibers, failing to fully exploit the potential of fiber reinforcement in mechanical strength.

\item \textbf{Turbulence:} In regions with stress concentration, vectors tend to exhibit disordered or chaotic directions (see Fig.~\ref{fig:Turbulence} for an example). This sort of turbulence leads to tremendous difficulty in maintaining the consistency required for effective fiber alignment along with the direction parallel to maximal principal stresses. 
\end{itemize}

Furthermore, planar layer based toolpath generation algorithms used in commercial 3D printing systems for CFRTP composites cannot effectively align fibers along with the maximal principal stresses that are distributed in three-dimensional space. This misalignment significantly reduces the effectiveness of fiber reinforcement. Differently, \rev{multi-axis additive manufacturing (MAAM)}{non-planar toolpaths} enables dynamic adjustment of the deposition direction and position during the printing process, where this flexibility allows the fiber deposition to be extended from traditional two-dimensional planes to three-dimensional spaces~\cite{Liu_SIG24}. When fibers are aligned with the maximal principal stress directions in 3D space, the loads are efficiently transferred to the fibers, thereby significantly enhancing the mechanical strength of a CFRTP composite part.

\subsection{Our Method}
In this paper, we present a toolpath generation method for high density fiber placement in CFRTP composites by using multi-axis additive manufacturing. The curved layer generation of our method is based on $S^3$-Slicer~\cite{Zhang_SIGA22} under the guidance of an input stress field with an extension to handle the winding compatibility around holes. To eliminate the sparsity of fiber toolpaths in prior approaches while allowing fibers to follow the indicated maximal stress directions, we make the following technical contributions in this paper to achieve nearly equal hatching distance between toolpaths (therefore also high density fiber placement).

\begin{itemize}
\item A 2-RoSy representation is introduced for formulating and computing the optimized direction field so that eliminates the influence of ambiguous orientation and aligns well with maximal principal stresses in the field-guided toolpath generation (Sec.~\ref{subsecDirFieldOptm}).

\item A method is developed to convert the direction field in 2-RoSy representation into a periodic scalar field that can help generate partial iso-curves for fiber toolpaths with nearly equal hatching distance (Sec.~\ref{subsecPeriodicFieldStripeToolpath}).

\item The $S^3$-Slicer is extended to account for winding compatibility around holes to improve fiber coverage in stress-concentrated regions (Sec.~\ref{secHoleCase}).
\end{itemize}
Our method provides a universal solution for generating continuous fiber toolpaths with nearly equal distance in 3D space, achieving good alignment between the stress field and fiber distribution. 

To verify the effectiveness of fiber toolpaths generated by our method, a new printer head with the capability to extrude matrix materials (PLA), support materials (PVA), and continuous carbon fiber (Markforged CF-FR-50) has been designed and integrated into our ABB robotic system that includes a 6-DOF robot arm and a 2-DOF positioner. Details of our hardware will be provided in Sec.~\ref{subsecHardware}.

Mechanical tests have been conducted to evaluate the mechanical strength of CFRTP composites fabricated by using our toolpath. Images of Scanning Electron Microscopy (SEM) have been taken to further analyze the failure of CFRTP composites and verify the mechanical strength. 
\section{Curved Slicer Based on Deformation}\label{secCurvedSlicer}

Our toolpath generation method presented in this paper adopts the deformation based $S^3$-Slicer~\cite{Zhang_SIGA22} to generate curved layers for CFRTP composites. As a preliminary step of toolpath generation, we provide a brief introduction to the basic concept and formulation of the $S^3$-Slicer below.
\begin{figure}[t]
\centering  
\includegraphics[width=0.75\linewidth]{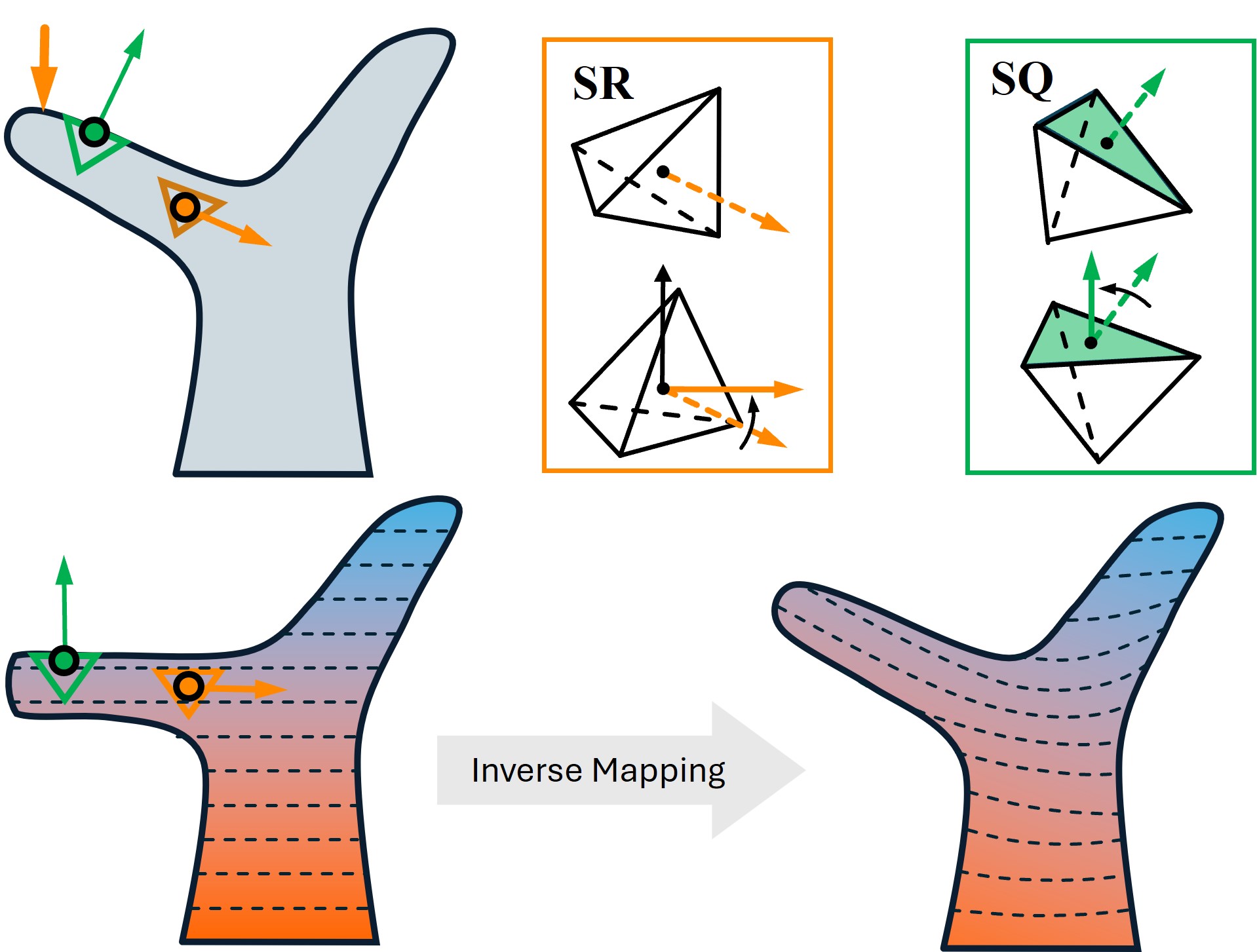}\vspace{-5pt}
\put(-200,76){\footnotesize \color{black}(a)}
\put(-126,76){\footnotesize \color{black}(b1)}
\put(-61.5,76){\footnotesize \color{black}(b2)}
\put(-200,3){\footnotesize \color{black}(c)}
\put(-80,3){\footnotesize \color{black}(d)}
\put(-180,20){\footnotesize \color{black}$\mathcal{M}^d$}
\put(-180,96){\footnotesize \color{black}$\mathcal{M}$}
\put(-60,20){\footnotesize \color{black}$\mathcal{M}$}
\put(-197,135){\footnotesize \color{black}$F$}
\put(-170,131){\footnotesize \color{black}$\mathbf{n}_f$}
\put(-160,100){\footnotesize \color{black}$\sigma_{\max}$}
\put(-103,100){\footnotesize \color{black}$\mathbf{d}_p$}
\put(-40,102){\footnotesize \color{black}$\mathbf{d}_p$}
\put(-10,135){\footnotesize \color{black}$\mathbf{n}_f$}
\caption{An illustration to explain the deformation-based $S^3$-Slicer~\cite{Zhang_SIGA22}: 
(a) The initial input volume model $\mathcal{M}$ under applied force $F$, where the maximal principal stress $\sigma_{\text{max}}$ in an element and the normal vector $n_f$ at a vertex on the boundary surface are also illustrated; Tetrahedral elements are applied with local rotation for enhancing mechanical strength (SR) and improving surface quality (SQ). The local printing directions $\mathbf{d}_p$ (as $z$-axis in the deformed shape) are made to be perpendicular to the maximal stress (b1) or be parallel to the surface normal (b2) by rotating every element. (c) The deformed model $\mathcal{M}^d$ is obtained through a scale-controlled deformation algorithm that stitches locally rotated elements together. The resulting field meets both the SR and the SQ requirements.
(d) The height field of a deformed model is mapped back to the original model $\mathcal{M}$ as a scalar field so that the curved layers can be extracted as the iso-surfaces of the scalar field.
}\label{fig:QDeformationSlicing}
\vspace{-5pt}
\end{figure}

\begin{figure*}[t]
\centering  
\includegraphics[width=\linewidth]{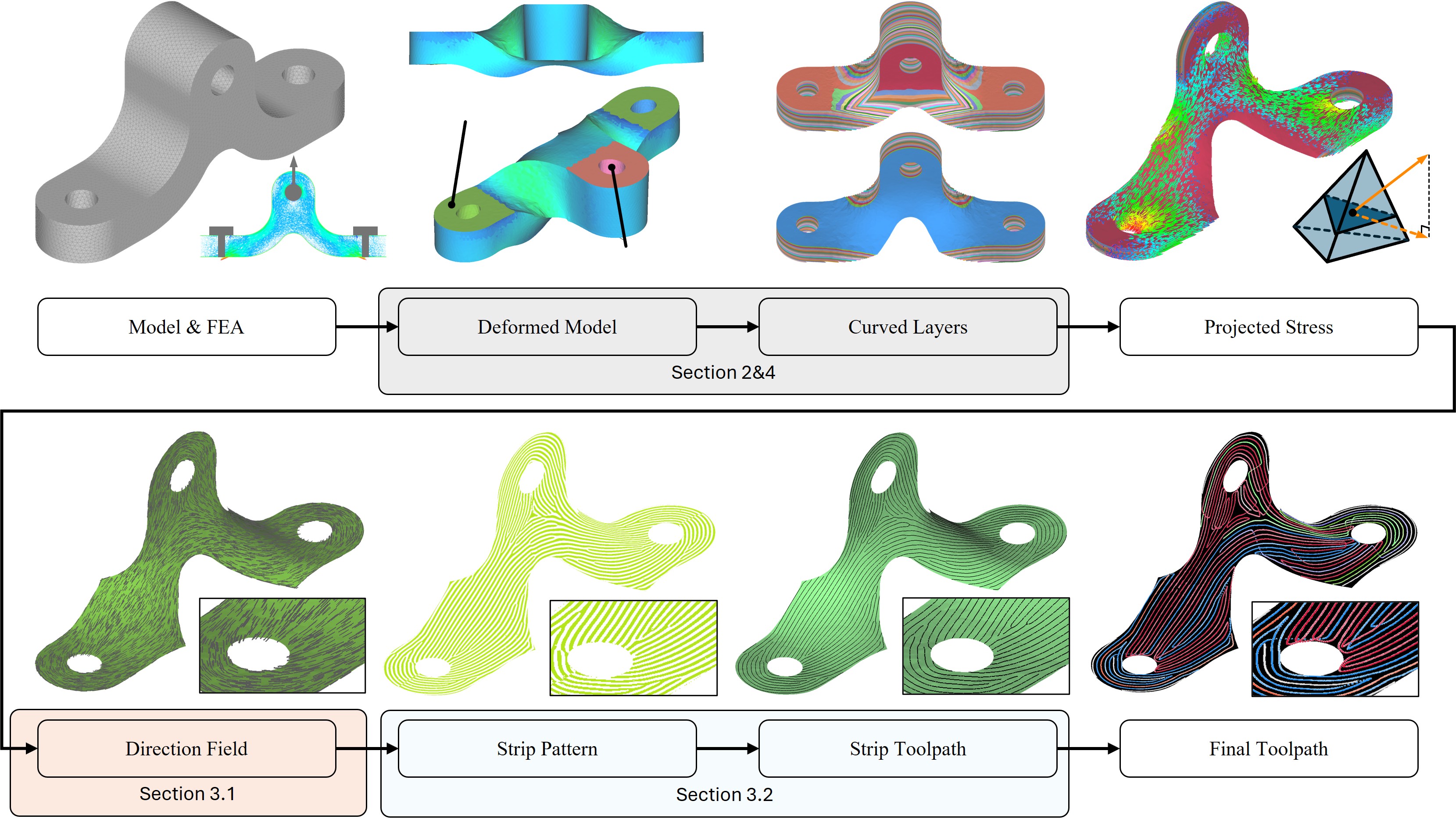}\vspace{-5pt}
\put(-530,290){\footnotesize \color{black}(a)}
\put(-400,290){\footnotesize \color{black}(b)}
\put(-265,290){\footnotesize \color{black}(c)}
\put(-130,290){\footnotesize \color{black}(d)}
\put(-530,140){\footnotesize \color{black}(e)}
\put(-400,140){\footnotesize \color{black}(f)}
\put(-265,137){\footnotesize \color{black}(g)}
\put(-130,137){\footnotesize \color{black}(h)}
\put(-390,263){\footnotesize \color{black}Region-of-Interest}
\put(-320,200){\footnotesize \color{black}Hole regions}
\put(-60,228){\footnotesize \color{black}$e$}
\put(-16,207){\footnotesize \color{black}$\sigma_{max}$}
\put(-20,249){\footnotesize \color{black}$\bar{\sigma}_{max}$}
\caption{The computational pipeline \rev{}{with steps as a diagram explanation of our method} for generating high density toolpaths for fiber placement: (a) the input T-bracket model is represented as a volumetric mesh $\mathcal{M}$ and the field of maximal principal stresses can be obtained from FEA by using the given loads as boundary conditions; (b) the deformed model $\mathcal{M}^d$ and (c) the curved layers are generated by the extension version of $S^3$-Slicer~\cite{Zhang_SIGA22}; (d) the field of maximal principal stresses $\sigma_{max}$ is projected from elements onto the curved layers (represented as triangular mesh surfaces); (e) the direction field $\mathbf{d}(\cdot)$ generated by our method following the stress field by using the 2-Rotational Symmetry (2-RoSy) representation, which means that there are two equivalent directions that are $180^{\circ}$ apart at every point; (f) the strip pattern obtained from the optimized periodic scalar field, which uses the $\mathbf{d}(\cdot)$ as input; 
(g) the toolpath generated from strip pattern by using the marching square algorithm~\cite{Maple_ICGMG03}; (h) the final toolpath for fiber placement after tracing connected toolpaths and removing the toolpaths shorter than a hardware-constrained length $\bar{L}$ that cannot be physically realized. \rev{}{Note that corresponding sections for different steps of the pipeline have been indicated on the flow chart.}
}\label{fig:pipeline}
\vspace{-5pt}
\end{figure*}
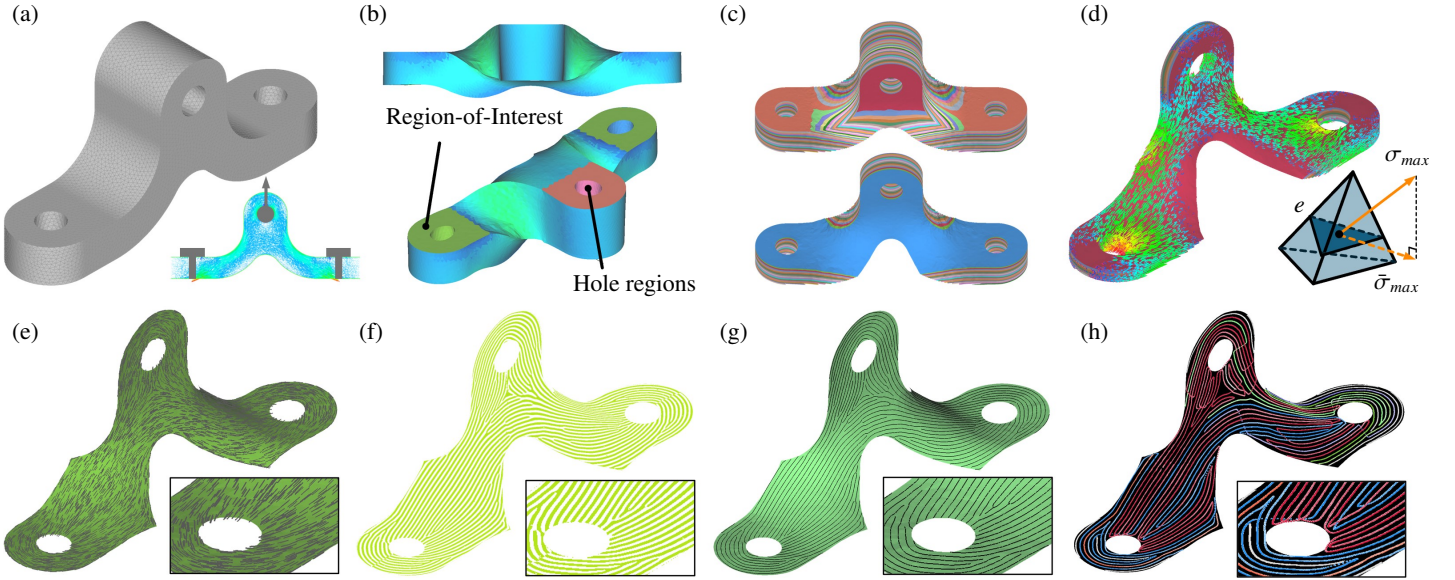

$S^3$-Slicer is a deformation-based optimization framework developed
to achieve multiple manufacturing objectives simultaneously. In this study, we focus on two primary objectives: the \textit{strength reinforcement} (SR) and the \textit{surface quality improvement} (SQ). The \textit{support-free} (SF) objective is omitted as the solvable supporting structures can be employed when fabricating CFRTP composites. Our aim in this toolpath generation research for CFRTP composites is to maximize the structural strength without significant compromise. As a curved slicer for multi-axis additive manufacturing, $S^3$-Slicer attempts to determine the optimized local printing directions (LPDs) in every element of an input model. The optimized rotation of each LPD is commonly determined by multiple objectives, subject to manufacturing constraints. Specifically, the input to the $S^3$-Slicer consists of the target model, which is discretized into a volumetric mesh $\mathcal{M}$ composed of a set of tetrahedral elements $\{e\}$. Another input, the stress field $\sigma_{\text{max}}$ under a given force $F$ is as illustrated in Fig.~\ref{fig:QDeformationSlicing}(a), which can be obtained from FEA.

The scalar field for generating curved layers is obtained by deforming the input model with the deformation driven by local rotations, which are represented by a field of quaternions~\cite{Zhang_SIGA22}. First, for the SR objective, the tetrahedral elements of the input model are rotated such that the maximal principal stresses in the deformed space become perpendicular to the printing direction (i.e., $z$-axis), ensuring that the maximal principal stress is aligned with the curved layers (see Fig.~\ref{fig:QDeformationSlicing}(b1)). Second, for the SQ objective, we define the region-of-interest (ROI) as the fixing and holding regions and ensure that these regions remain completely within a single layer, facilitating part mounting and load application. The normal direction of the boundary facets is aligned with the printing direction in the deformed space by rotating the tetrahedral elements accordingly (see Fig.~\ref{fig:QDeformationSlicing}(b2) for an illustration).

Besides of these local rotational operations described by quaternions, we also introduce an energy term to control the harmonics of neighboring rotations (i.e., quaternions) to prevent excessive rotations leading to unmanufacturable layers \rev{}{having high surface curvature, the fabrication of which will lead to local collision between the printer head and the working surface}. In short, the optimized field of quaternions can be computed by solving the following minimization problem:
\begin{equation}\label{eq:Local}
    \begin{aligned}
    \arg \min_{\{\mathbf{q}_e\}} \ \ & \underbrace{\sum_{e \in \mathcal{M}} w_{sr}(e) \| \mathbf{q}_e - \mathbf{q}_e^{tr} \|^2}_{\mathrm{Strength~Reinforcement} \; E_{sr}(\mathcal{M})}  + \underbrace{w_{sq} \sum_{e \in ROI} \| \mathbf{q}_e - \mathbf{q}_e^{tq} \|^2}_{\mathrm{Surface~Quality} \; E_{sq}(\mathcal{M})} \\ 
    & + \underbrace{\sum_{(e_i,e_j) \in \mathcal{N}_{f}} \| \mathbf{q}_{e_i} - \mathbf{q}_{e_j} \|^2}_{\mathrm{Harmonic~of~Field} \; E_{hf}(\mathcal{M})},
    \end{aligned}
\end{equation}
where $e_i$ and $e_j$ are a pair of face-adjacent elements in the set of neighboring elements $\mathcal{N}_{f}$. The first two terms are used to control the elements to follow the requirements of SR and SQ, and the last term is to ensure the smoothness of the quaternion field. The variable $\mathbf{q}_e$ denotes the quaternion associated with each element, while $\mathbf{q}_e^{tr}$ and $\mathbf{q}_e^{tq}$ represent the target rotations determined during the local projection steps for SR and SQ respectively. The value of $w_{sr}(e) \in [0,1]$ is chosen according to the value of maximal principal stress in each element by mapping the range of maximal stresses within the whole model to $[0,1]$. \rev{}{The value of $w_{sq}$ is determined by experimental tests and} $w_{sq}= 10.0$ is adopted in our implementation for all models.

In the model deformation step, all elements $\{e'\}$ of the input model are deformed according to the rotations determined from the above optimized quaternion fields and stitched together. The desired rotation on every element $e$ is represented as $\mathbf{R}(\mathbf{q}_e)(\mathbf{NV}_e)^\mathrm{T}$ with $\mathbf{V}_e$ being a position matrix formed by the coordinates of $e$'s vertices on the input model$\mathcal{M}$. The positions of vertices on the deformed model are stored in $\mathbf{V}_e^d$ and determined by solving the following optimization problem.
\begin{equation}\label{eq:deformASAP}
\begin{aligned}
\arg \min_{\mathcal{M}^d} & \ \ \underbrace{\sum_{e\in\mathcal{M}} \|(\mathbf{NV}_e^d)^\mathrm{T} - \mathbf{R}_e \mathbf{S}_e (\mathbf{NV}_e)^\mathrm{T} \|^2_F}_{\mathrm{Position~Compatibility}~E_{pc}(\mathcal{M}^d)} \\
& + \underbrace{w_{rs}\sum_{e\in\mathcal{M}}  \|\mathbf{S}_e - \mathbf{I} \|^2}_{\mathrm{Scale~Rigidity}~E_{rs}(\mathcal{M}^d)}
+ \underbrace{w_{sc}\sum_{(e_i,e_j) \in \mathcal{N}_f}  \|\mathbf{S}_{e_i} - \mathbf{S}_{e_j} \|^2 }_{\mathrm{Scale~Compatibility}~E_{sc}(\mathcal{M}^d)},
\end{aligned}
\end{equation}
\rev{where $\mathbf{N}$ transfers the element’s center to the origin as $\mathbf{N}=\mathbf{I}_{4\times4}-\frac{1}{4}\mathbf{1}_{4\times4}$. The scaling variables are introduced for each element as $\mathbf{S}_e = \mathrm{diag}(s_e^x,s_e^y,s_e^z)$ -- this gives a locally scaled and rotated element as $\mathbf{R}(\mathbf{q}_e)\mathbf{S}_e(\mathbf{NV}_e)^\mathrm{T}$. }{where $\mathbf{N}$ shifts the element’s center to the origin (refer to~\cite{Zhang_SIGA22}), defined as $\mathbf{N}=\mathbf{I}_{4\times4}-\frac{1}{4}\mathbf{1}_{4\times4}$. Scaling variables for each element are introduced as $\mathbf{S}_e = \mathrm{diag}(s_e^x,s_e^y,s_e^z)$, resulting in a locally scaled and rotated element represented by $\mathbf{R}(\mathbf{q}_e)\mathbf{S}_e(\mathbf{NV}_e)^\mathrm{T}$.}
In the second term, we control the rigidity of scaling by measuring the difference between $\mathbf{S}_e$ and $\mathbf{I}$. The compatibility of scales between neighboring elements is controlled by adding the third term. In our implementation, $w_{rs}=1.0$  and $w_{sc}=3.0$ are always chosen \rev{}{through computational experiments} to balance the importance of different terms. \rev{}{Note that Eqs.\eqref{eq:Local} and \eqref{eq:deformASAP} are both in the least-squares form and can be computed by solving a linear equation system.}

After applying the above rotation-driven deformation, the height field $H^d(\textbf{x})$ of the deformed model $\mathcal{M}^d$ (see Fig.~\ref{fig:QDeformationSlicing}(c)) is mapped back to become a scalar field $G(\textbf{x})$ with $\textbf{x} \in \mathbb{R}^3$ defined on the input model, from which curved layers for multi-axis additive manufacturing are extracted as iso-surfaces of the scalar field $G(\textbf{x})$ (Fig.~\ref{fig:QDeformationSlicing}(d)). 
As illustrated in Fig.~\ref{fig:pipeline}(a-c), the extended $S^3$-Slicer is employed in our computational pipeline to generate the curved layers for the additive manufacturing of CFRTP composites. Details of $S^3$-Slicer extension will be presented in Sec.~\ref{secHoleCase}.

\section{Toolpath Generation with Controlled Hatching Distance}\label{secStripToolpath}

After obtaining the curved layers as mesh surfaces and projecting the field of maximal principal stresses onto the meshes, we have stress information for the toolpath generation. In order to overcome the challenges discussed in Sec.~\ref{subSec:ProblemState}, we first generate a direction field $\mathbf{d}(\cdot)$ following the projected maximal principal stresses and then convert it into a periodic scalar field $G(\cdot)$ with its gradient perpendicular to $\mathbf{d}(\cdot)$ -- i.e., the iso-curve (toolpath) of $G(\cdot)$ will follow the direction field $\mathbf{d}(\cdot)$. Specifically, the 2-RoSy representation is employed to compute the direction field in Sec.~\ref{subsecDirFieldOptm} so that the problems of directional ambiguity and turbulence can be solved. The scalar field $G(\cdot)$ is parameterized by using a spinning form in Sec.~\ref{subsecPeriodicFieldStripeToolpath} so that iso-curves with nearly equal hatching distance can be extracted from $G(\cdot)$ for toolpaths, achieving high density fiber coverage.

\subsection{Optimization of Direction Field}\label{subsecDirFieldOptm}
\subsubsection{Preparation: 2-RoSy representation} 

In the previous work (e.g.,~\cite{Fang_ADDMA24,Fang_SIGA20}), the toolpath of fibers is generated by the iso-curves of a scalar field with its gradients defined by a vector field. However, the conventional vector field based representation does not consider the directional symmetry of fiber placement. Specifically, given the maximal stress $\sigma_{\max}$, aligning fibers along the directions $\sigma_{\max}$ or $-\sigma_{\max}$ will introduce the same reinforcement. A new representation treating the vectors $\pm \mathbf{v}$ equally is needed.

To address this issue, we borrow the 2-Rotationally Symmetric (2-RoSy) representation in this paper to compute the direction field, which is characterized by its 180-degree rotational symmetry. Considering a unit vector $\mathbf{v} = \cos \theta \, \mathbf{u}_1 + \sin \theta \, \mathbf{u}_2$ with $\mathbf{u}_1$ and $\mathbf{u}_2$ being the basis vectors of a local frame, we can convert all vectors $\mathbf{v}$ into a 2-RoSy representation $\mathbf{d}$ as:
\begin{equation}
    \mathbf{d} = \mathbf{d}(\theta) = \begin{bmatrix} \cos(2\theta) \\ \sin(2\theta) \end{bmatrix},
    \label{eqn:rosy}
\end{equation}
All processes on a vector field $\mathbf{v}(\cdot)$ requiring the directional symmetry can then be processed on the field $\mathbf{d}(\cdot)$ instead of $\mathbf{v}(\cdot)$. Afterwards, the resultant $\mathbf{d}(\cdot)$ can be easily converted back into $\mathbf{v}(\cdot)$ when needed.

It is straightforward to verify that this representation is directional symmetry -- i.e., invariant under a 180-degree rotation of the original vector, as $\mathbf{d}(\theta \pm \pi) = \begin{bmatrix} \cos(2\theta \pm 2\pi) \\ \sin(2\theta \pm 2\pi) \end{bmatrix} = \mathbf{d}(\theta)$. 

\subsubsection{Assignment of local frames}
The toolpath for fiber placement is computed on curved layers that are generated from $S^3$-Slicer and stored in the discrete form as $\{\mathcal{S}^i\}$ where each layer is
denoted by $\mathcal{S} = \{\mathcal{V}, \mathcal{E}, \mathcal{F}\}$ with $\mathcal{V}$, $\mathcal{E}$ and $\mathcal{F}$ representing the sets of vertices, edges and faces. For governing the toolpath generation aligned with the projected maximal stress $\bar{\sigma}_{\max}$, we convert each $\bar{\sigma}_{\max}(f)$ on a face $f$ into a 2-RoSy vector $\mathbf{s}_f$ for generating the direction field $\mathbf{d}(\cdot)$. The direction field will be represented as a piece-wise linear function of 2-RoSy vectors with $\mathbf{d}_{v_i}$ being defined on each vertex $v_i$ of the triangular mesh $\mathcal{S}$.

Note that in the toolpath generation, the direction field is defined on the 2D manifold, which is a non-flat space. Therefore, the definition of direction similarity not only needs to consider the directions themselves but also needs to consider the differences between neighboring local coordinate systems in which the directions are located. According to the theory of discrete differential geometry\rev{}{~\cite{Crane_CGF10, Knoppel_book16, Knoppel_SIG15}} about the construction of the local coordinate systems, we build polar coordinates at each vertex $v_{i} \in \mathcal{V}$. \rev{The direction vector $\mathbf{d}_{v_i}$ is treated as angles $\theta_i \in [-\pi, \pi)$ relative to the first edge $e_{ij_{0}} \in \mathcal{E}$ chosen at each vertex instead of viewing as elements of $\mathbb{R}^3$. This intrinsic setup simplifies the rest of the algorithm and eliminates the need to define normals at vertices.}{ The direction vector $\mathbf{d}_{v_i}$ is represented intrinsically as an angle $\theta_i \in [-\pi, \pi)$ measured relative to the first edge $e_{ij_{0}} \in \mathcal{E}$ selected at each vertex, rather than being treated as a vector in $\mathbb{R}^3$. This approach streamlines the algorithm by avoiding the requirement to define vertex normals and simplifies the subsequent computations.}

\begin{figure}[t]
\centering
\includegraphics[width=\linewidth]{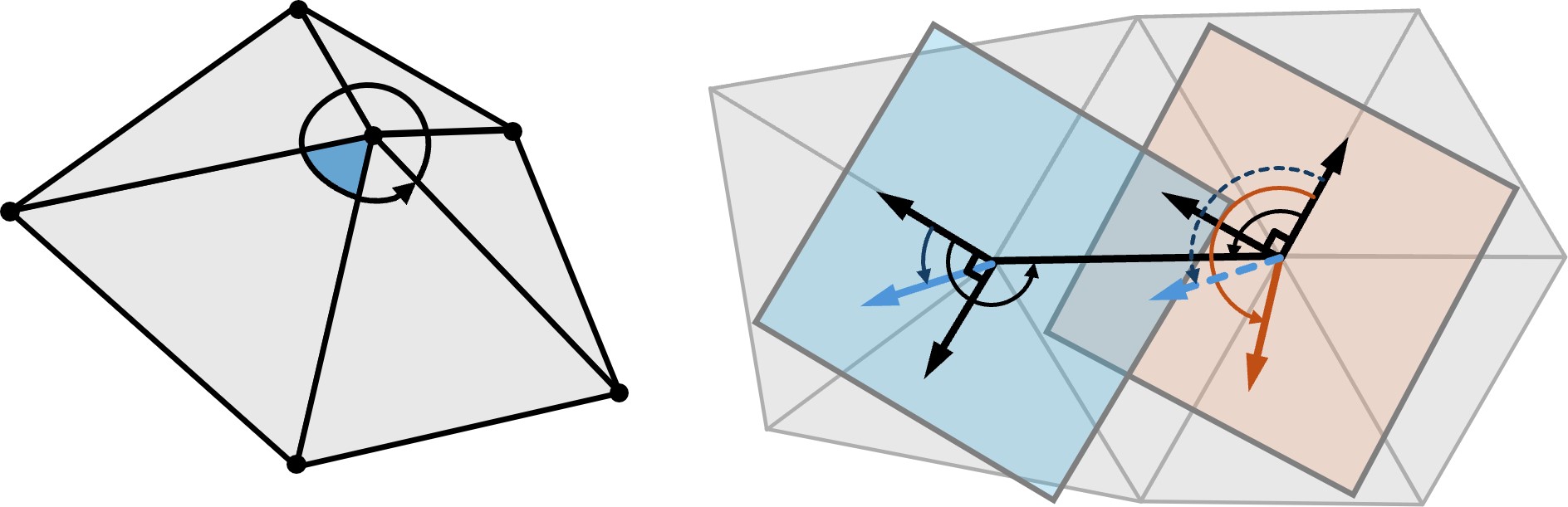}
\put(-257,0){\footnotesize \color{black}(a)}
\put(-144,0){\footnotesize \color{black}(b)}
\put(-190,75){\footnotesize \color{black}$i$}
\put(-154,12){\footnotesize \color{black}$j_0$}
\put(-243,44){\footnotesize \color{black}$j_p$}
\put(-205,16){\footnotesize \color{black}$j_{p+1}$}
\put(-217,64){\footnotesize \color{black}$\Theta_i$}
\put(-222,42){\footnotesize \color{black}$\hat{\theta}_i^{j_{p}j_{p+1}}$}
\put(-96,44){\footnotesize \color{black}$i$}
\put(-45,33){\footnotesize \color{black}$j$}
\put(-110,52){\footnotesize \color{black}$\mathbf{u}_1$}
\put(-100,20){\footnotesize \color{black}$\mathbf{u}_2$}
\put(-50,60){\footnotesize \color{black}$\mathbf{u}_1$}
\put(-75,50){\footnotesize \color{black}$\mathbf{u}_2$}
\put(-90,30){\footnotesize \color{black}$\theta_{ij}$}
\put(-43,43){\footnotesize \color{black}$\theta_{ji}$}
\put(-120,39){\footnotesize \color{black}$\phi_i$}
\put(-123,29){\footnotesize \color{black}$\mathbf{d}_{v_i}$}
\put(-74,32){\footnotesize \color{black}$\phi_i^{\prime}$}
\put(-61,25){\footnotesize \color{black}$\phi_j$}
\caption{The definition and application illustration of the local frame on a 2D manifold. (a) Polar coordinates defined at each vertex $v_i \in \mathcal{V}$. (b) Illustration of the method for measuring the angle between two adjacent vectors on the manifold. Note that the unit vector $\mathbf{d}_{v_i}$ and the angle $\phi_i$ are equivalent here, representing the same quantity in different coordinate systems.}
\label{fig:localFrame}
\end{figure}
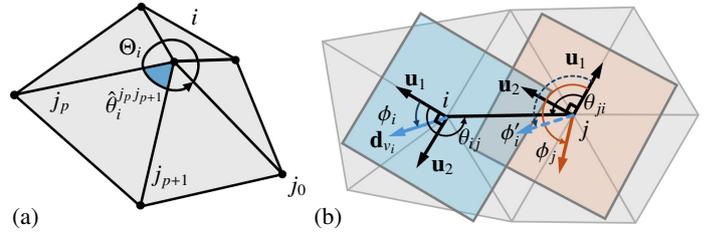

Moreover, let $\hat{\theta}_i^{jk}$ denote the angle at corner $i$ of the triangle $f_{ijk} \in \mathcal{F}$. \rev{$\Theta_i := \sum_{f_{ijk} \in \mathcal{F}} \hat{\theta}_i^{jk}$ can be the sum of all angles incident on vertex $v_i$ (see Fig.~\ref{fig:localFrame}(a) for an illustration).}
{The total angle around vertex $v_i$ is given by $\Theta_i := \sum_{f_{ijk} \in \mathcal{F}} \hat{\theta}_i^{jk}$, summing all angles associated with $v_i$ (see Fig.~\ref{fig:localFrame}(a) for an example).} For each oriented edge $e_{ij_{a}}$ incident on $i$, the angle is defined as

\begin{equation}
    \theta_{ij_a} := \frac{2\pi}{\Theta_i} \sum_{p=0}^{a-1} \hat{\theta}_i^{j_{p}j_{p+1}},
\end{equation}
\rev{where $j_p$ is the counterclockwise index centered at vertex $i$ and $a$ is the degree of vertex $i$. These scaled angles effectively act as polar coordinates for the edge, starting from $\theta_{ij_0} = 0$, which means that the vector $\mathbf{u}_1$ follows the edge $e_{ij_0}$. }{where $j_p$ denotes the counterclockwise index around vertex $i$, and $a$ represents the degree of vertex $i$. These scaled angles serve as a form of polar coordinates for the edges, with $\theta_{ij_0} = 0$ indicating that the vector $\mathbf{u}_1$ aligns with the direction of the edge $e_{ij_0}$.}

\subsubsection{Metric of vector fields on a manifold}
After defining the local frames, we need to formulate the metric to evaluate the difference of vectors defined on different local frames. \rev{Based on the local frame defined before, any unit vector $\mathbf{d}_{v_i}$ can be encoded by the angle $\phi_i$, the parallel vector at the adjacent vertex $j$ can be computed by}{Using the previously defined local frame, any unit vector $\mathbf{d}_{v_i}$ can be described by the angle $\phi_i$. The parallel vector at an adjacent vertex $j$ is then computed as}
\begin{equation}
    \phi_i^{\prime} = \phi_i \, \underbrace{ - \theta_{ij} + (\theta_{ij} + \pi)}_{\rho_{ij}},
\end{equation}
using the shared edge $e_{ij}$ as a common frame of reference (see Fig.~\ref{fig:localFrame}(b)). Here we store the values $\rho_{ij} := - \theta_{ij} + (\theta_{ji} + \pi)$, noting that $\rho_{ji} = -\rho_{ij}$. Based on this, the coordinate transformation matrix $\mathbf{R}_{ij}$ between two neighboring local frames in vertices ${i}$ and ${j}$ can be defined as:
\begin{equation}\label{eq:transformMatrix}
    \mathbf{R}_{ij} := \begin{bmatrix}
                \cos (\rho_{ij}) & -\sin (\rho_{ij}) \\
                \sin (\rho_{ij}) & \cos (\rho_{ij})
             \end{bmatrix}.
\end{equation}
With the help of $\mathbf{R}_{ij}$, two adjacent vectors on the 2D manifold can be moved to the same local coordinate system, so that the difference (distance) between the vectors can be effectively defined as $\| \mathbf{d}_{v_i} - \mathbf{R}_{ij} \mathbf{d}_{v_j} \|$ and the optimization of the direction field becomes possible.

\subsubsection{Local constraints and optimization}
The toolpath planning for spatial fiber placement on each layer requires adherence to the following two criteria:
\begin{enumerate}
\item To maximize the stiffness of parts printed with anisotropic materials (continuous carbon fiber), the maximal principal stress direction has proven to be an optimal fiber orientation~\cite{Suzuki_CMAME91}. 
\item The field exhibits significant directional variations in regions with turbulent. Then toolpaths extracted in these areas will lead to fiber fabrication failure \rev{}{(i.e., sharp turns of toolpaths on a curved layer)}. Therefore, the harmonic requirement is usually introduced to guide the orientation of fiber placement in such regions.
\end{enumerate}
Both these criteria will be met by first computing an optimized direction field $\mathbf{d}(\cdot)$ for supervising the toolpath generation. The direction field satisfying the aforementioned criteria can be computed by solving the following optimization problem:
\begin{equation}\label{eq:directionFieldOpt}
    \arg \min_{\mathbf{d}_{v_i}}
    \underbrace{
    \frac{1}{|\mathcal{V}|}\sum_{v_{i}\in \mathcal{V}} \frac{1}{|\mathcal{N}(v_i)|}\sum_{f\in \mathcal{N}(v_i)} \| \mathbf{d}_{v_i} - \mathbf{s}_{f} \|^2
    }_{\mathrm{Stress~Alignment} \; E_{sa}(\mathcal{S})}
    + \underbrace{\frac{\omega_k}{|\mathcal{E}|}\sum_{e_{ij}\in \mathcal{E}} \| \mathbf{d}_{v_i} - \mathbf{R}_{ij} \mathbf{d}_{v_j} \|^2}_{\mathrm{Harmonic~of~Field} \; E_{uf}(\mathcal{S}) \;  }, 
\end{equation}
where $|\cdot|$ denotes the number of elements in a set, $\mathbf{d}_{v_i}$s are undetermined variables as 2-RoSy vectors defined on vertices $\{v_i\}$ and $\mathbf{s}_{f}$ are 2-RoSy representation of the projected maximal stresses as discussed above. In the second term, $\mathbf{R}_{ij}$ is the rotation matrix that aligns the local frame at vertex $j$ to the local frame at vertex $i$ (refer to Eq.~\eqref{eq:transformMatrix}). This minimization problem is in the form of least-squares, and $\omega_k$ is a weight balancing the two objectives. We use $\omega_k = 4.0$ for all examples presented in this paper.

The solution to this least-squares problem can be computed through an equivalent linear system:
\begin{equation}
    \begin{bmatrix}
        \mathbf{B}^s \\
        \mathbf{B}^h
    \end{bmatrix}
    \mathbf{d} = 
    \begin{bmatrix}
        \mathbf{s} \\
        \mathbf{0}
    \end{bmatrix}
    \label{eqn:linearSystem}
\end{equation}
where $\mathbf{B}^s \in \mathbb{R}^{2|\mathcal{V}|\times 2|\mathcal{V}|}$ is a diagonal matrix derived from the stress-alignment term, and $\mathbf{B}^h \in \mathbb{R}^{2|\mathcal{V}|\times 2|\mathcal{V}|}$ is a square matrix as follows:
\begin{equation}
    \mathbf{B}^h_{[2i,2j]} = \left\{
    \begin{aligned}
        & -\mathbf{R}_{ij},  & \text{if } e_{ij} \in \mathcal{E}, \\
        & \mathbf{I}_{2\times 2}, & \text{if } i = j, \\
        & \mathbf{0}, & \text{otherwise}.
    \end{aligned}
    \right.
\end{equation}
Here $\mathbf{B}^h_{[2i,2j]}$ represents the sub-matrix of $\mathbf{B}^h$ at rows of $2i, 2i+1$ and columns of $2j, 2j+1$. The matrix $\mathbf{R}_{ij}$ is the rotation matrix aligning the local frame at vertex $v_j$ to the local frame at vertex $v_i$ (Eq.~\eqref{eq:transformMatrix}), and $\mathbf{I}_{2\times 2}$ is a $2\times 2$ identity matrix.

The solution of this optimization will be a direction field $\mathbf{d}(\cdot)$
that adheres to the input constraints of stress directions while minimizing sudden variations, thereby enabling the continuous placement of fibers in multi-axis additive manufacturing.

\subsection{Stripe toolpath generation}\label{subsecPeriodicFieldStripeToolpath}
We now introduce the method to compute the scalar field $G(\cdot)$ whose gradient is perpendicular to a given direction field $\mathbf{d}(\cdot)$ and to generate the toolpath from $G(\cdot)$'s iso-curves to cover the surface $\mathcal{S}$ as completely as possible.

This problem can be viewed as a global parameterization challenge. Related works~\cite{Fang_ADDMA24, Chen_ADDMA22} implicitly represent the direction of the fiber toolpath using the iso-curves of a scalar field $G$, which provides a non-periodic parameterization. However, this method does not ensure the generation of a toolpath with equal hatching space  (see Fig.~\ref{fig:problem_statement}(b2\&b3) for an illustration of these issues). In contrast, inspired by~\cite{Knoppel_SIG15}, a periodic parameterization by spinning form can represent the vanishing or appearance of paths, resulting in a toolpath with nearly equal hatching space, as shown in Fig.~\ref{fig:problem_statement}(b1).

\subsubsection{Spinning form}
The spinning form is a periodic function over a triangular mesh at each vertex, $G: \mathcal{V} \rightarrow \mathcal{\psi}$, where the field value at each vertex $v_i$ is defined by a proxy representation as $G_i= G(v_i) = [\cos \psi_i \; \sin \psi_i]^T$.
The toolpath extraction is based on the iso-curves of the scalar field $G$ that follow the direction field $\mathbf{d}(\cdot)$. In order to achieve it, we control its gradient $\nabla G$ vertical to the given field $\mathbf{d}$ as $\nabla G \perp \mathbf{d}$, i.e., the iso-curve (toolpath) will follow the given smooth direction field.


Directly calculating the gradient $\nabla G$ at each vertex is challenging due to the incomplete and non-differentiable nature of the neighboring vertices. Alternatively, the gradient of $G(\cdot)$ at a vertex $v_i$ along its adjacent edge $e_{ij}$ can be computed by $(G_i - G_j)/\|e_{ij}\|$ -- denoted by $D_{e_{ij}} G(v_i)$. We then rotate the target direction $\mathbf{d}_i$ at $v_i$ anti-clockwise in $90^{\circ}$ to obtain $\mathbf{d}_i^{\perp}$. The angle between $\mathbf{d}_i^{\perp}$ and $e_{ij}$ is defined as $\phi_{ij}$. Based on these preparations, we convert the requirement of $\nabla G \perp \mathbf{d}$ into the objective of minimizing the difference between $D_{e_{ij}} G(v_i) \cos(\phi_{ij})$ and $\mathbf{d}_i^{\perp}$. This can be computed by solving the following optimization problem:

\begin{figure}[t]
\centering
\includegraphics[width=0.45\linewidth]{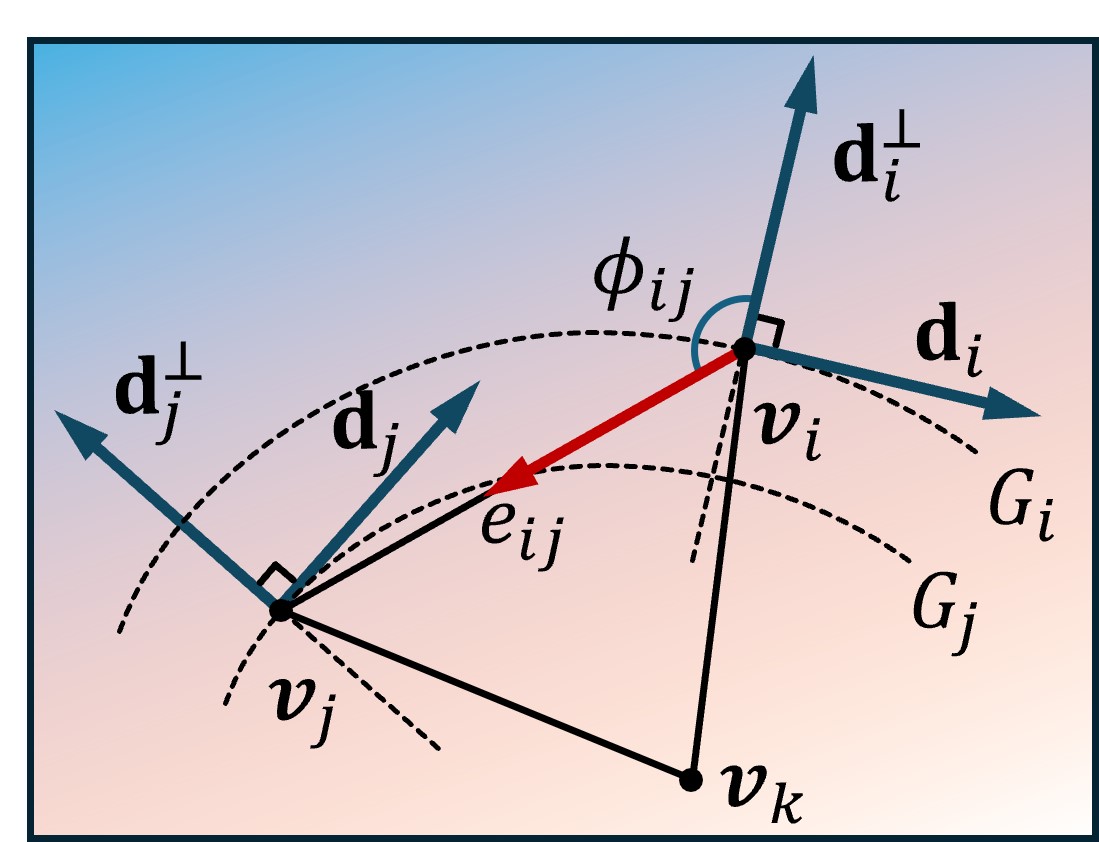}
\caption{Illustration of the annotation of a spinning form, represented as a periodic function applied to each vertex of a triangular mesh.}\label{fig:spiningForm}
\end{figure}

\begin{equation}\label{eq:stripForm}
        \arg \min_{G}
\frac{1}{|\mathcal{E}|} \sum_{e_{ij} \in \mathcal{E}}  \,
    \left\| \frac{\cos \phi_{ij}}{\|e_{ij}\|} (G_i - G_j) - 
    \mathbf{d}_i^{\perp} \right\|^2,
\end{equation}
that is
\begin{equation}\label{eq:stripForm2}
    \arg \min_{\psi}
\frac{1}{|\mathcal{E}|} \sum_{e_{ij} \in \mathcal{E}}  \,
    \left\| \frac{\cos \phi_{ij}}{\|e_{ij}\|} \begin{bmatrix} \cos \psi_i - \cos \psi_j \\ \sin \psi_i - \sin \psi_j\\ \end{bmatrix} - 
    \begin{bmatrix} -\sin 2\theta_i \\ \cos 2\theta_i\\ \end{bmatrix} \right\|^2
\end{equation}
with $\mathbf{d}_i=[\cos 2\theta_i \; \sin 2\theta_i]^T$ being determined from the direction field and then $\mathbf{d}_i^{\perp}=[-\sin 2\theta_i \; \cos 2\theta_i]^T$. The resultant field $G(\cdot)$ has both its $x$- and $y-$components defined as periodic functions. The iso-curves can be either extracted from $G^x(\cdot)$ or $G^y(\cdot)$, which have nearly similar distributions but only $\pi/2$ phase shift. \rev{}{To avoid the trivial solution -- all \textit{zeros}, this equation is solved by Eigen value decomposition.}

\begin{figure}[t]
\centering
\includegraphics[width=0.99\linewidth]{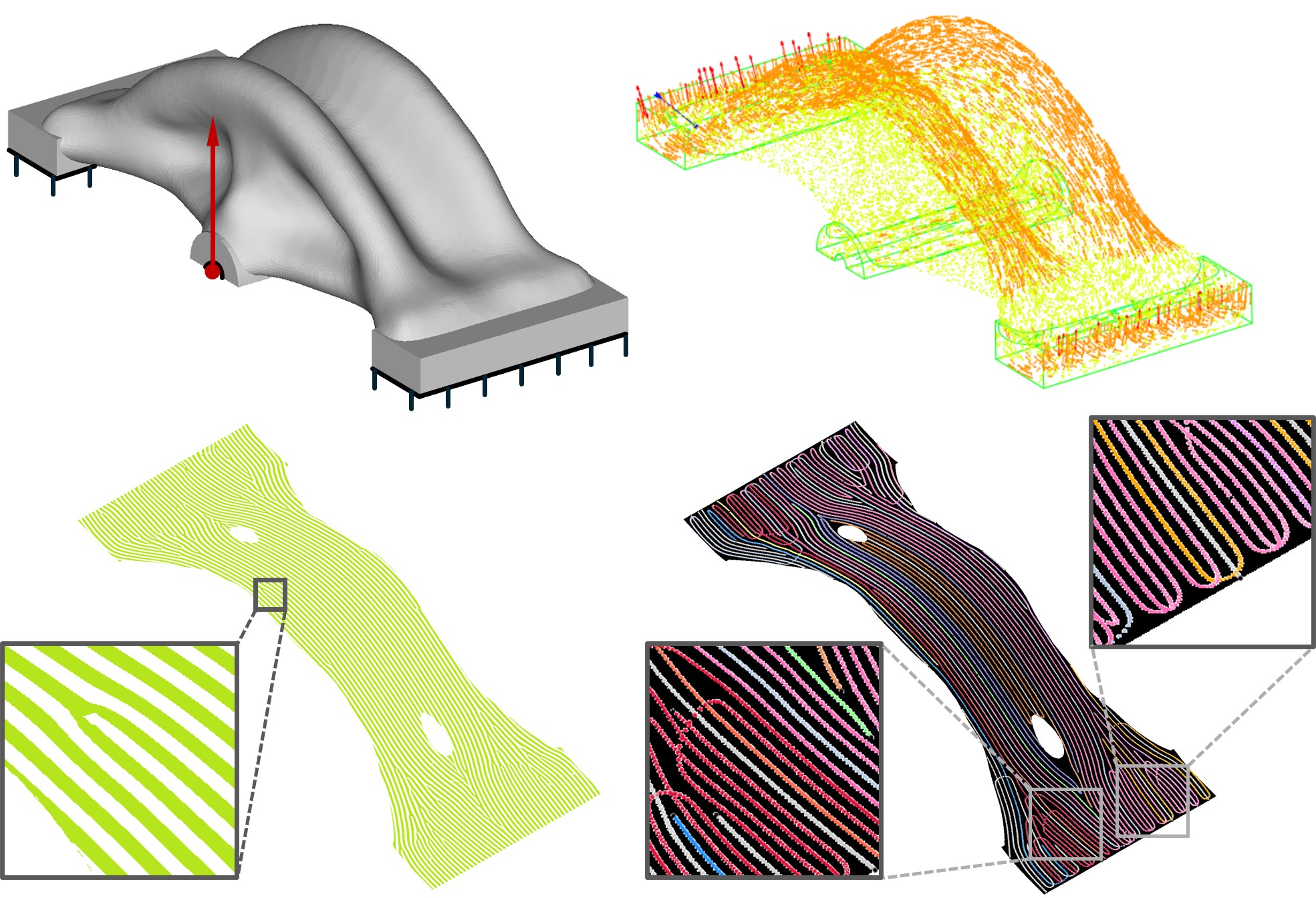}
\put(-259,173){\footnotesize \color{black}(a)}
\put(-240,120){\footnotesize \color{black}Force}
\put(-206,100){\footnotesize \color{black}Fixed}
\put(-130,173){\footnotesize \color{black}(b)}
\put(-259,85){\footnotesize \color{black}(c)}
\put(-130,85){\footnotesize \color{black}(d)}
\caption{The computational results for the Bridge model: (a) the input model and loads as boundary conditions; (b) the max. principal stress field under the given boundary conditions; (c) the strip pattern with nearly equal hatching spacing distance on a curved layer; (d) the fiber toolpaths extracted from the strip pattern that have been processed according to the manufacturing constraints. 
}\label{fig:bridgeRes}
\end{figure}

\subsubsection{Processing iso-curves}
After applying periodic parameterization in the spanning form to obtain the scalar field $G: \mathcal{V} \rightarrow \psi$ on a surface mesh, we can extract the iso-curves by using the Marching Squares algorithm~\cite{Maple_ICGMG03} from either $G^x(\cdot)$ or $G^y(\cdot)$.
It is important to note that singularities may occur at locations where strips appear or vanish due to inconsistencies in the number of strips flowing in and out (see the zoom-view of Fig.~\ref{fig:bridgeRes}(c)).
Authors of~\cite{Sharp_SIG19} introduced a sophisticated method to handle such cases. Considering the special property of fibers such as continuous carbon fibers with very high stiffness -- therefore hard to include sharp-turns in a toolpath~\cite{Huang_AM2023,Huang_TOG2024}, we employ a relatively simpler method to handle such cases. We basically skip the iso-curve generation in the triangles with such singularity and leave the task of toolpath generation in these triangles in the later step of path tracing. Details are given below.

The iso-curves generated by using the Marching Square algorithm are in the form of unsorted line segments in the triangles of the mesh surface $\mathcal{S}$. The line segments are sorted and connected together by a tracing algorithm, which is accelerated by using the connectivity information of $\mathcal{S}$. When tracing into a triangle having the aforementioned singularity, we connect the toolpath to the next nearest node located on the boundary of the triangle if this new line segment will not intersect with the already traced toolpath. 
\rev{}{Note that toolpath intersections occur only during transitions between different strip paths, and these intersections are limited to single points. Since there is no continuous overlap between paths, the impact of these intersections on mechanical performance is negligible.} Sharp-turns \rev{}{with curve radius less than 4~mm will lead to printing problems as analyzed in~\cite{Matsuzaki_ADDMA18}. They} are only allowed on a toolpath when near the boundary of $\mathcal{S}$, where a sharp-turn can be achieved by moving the print head outside the print area to access the next fiber path.
After tracing, the toolpaths that are shorter than the distance $\bar{L}$ between the fiber cutter and the print head are removed as they are not manufacturable. In our hardware setup, $\bar{L} = 70\mathrm{mm}$. 
\section{Winding Compatibility around Holes}\label{secHoleCase}
As discussed in our earlier work~\cite{Fang_ADDMA24}, laying fibers around mounting holes can improve model strength. The new approach proposed in this paper can also incorporate a similar consideration of winding compatibility around the hole with minor modifications. This is treated as an optional process in our framework.

First, in the deformation-based $S^3$-Slicer, the middle axis of the hole is expected to align with the printing direction in the deformation space, so that the layers with large areas can be formed -- i.e., better fiber winding. To reflect this requirement, a hole orientation term is added to enhance Eq.~\eqref{eq:Local} as
\begin{equation}\label{eq:Local_hole}
    \begin{aligned}
    \arg \min_{\{\mathbf{q}_e\}} & \ \ E_{sr}(\mathcal{M}) + E_{sq}(\mathcal{M}) + E_{hf}(\mathcal{M}) \\
    & + \underbrace{ w_{hw} \sum_{e \in \mathcal{H}}  \| \mathbf{q}_e - \mathbf{q}_e^{th} \|_2^2}_{\mathrm{Hole~Orientation}~E_{ho}(\mathcal{M})},
    \end{aligned}
\end{equation}
where $\mathbf{q}_e^{th} = (0, 1, 0)$ represents the orientation of holes in the deformation space, and $w_{hw}=100.0$ is employed as the weight to balance this target with other objectives. The set $\mathcal{H}$ is defined as a collection of tetrahedral elements that are face-adjacent to the holes with orientations to be controlled.

\begin{figure}[b]
\centering  
\includegraphics[width=\linewidth]{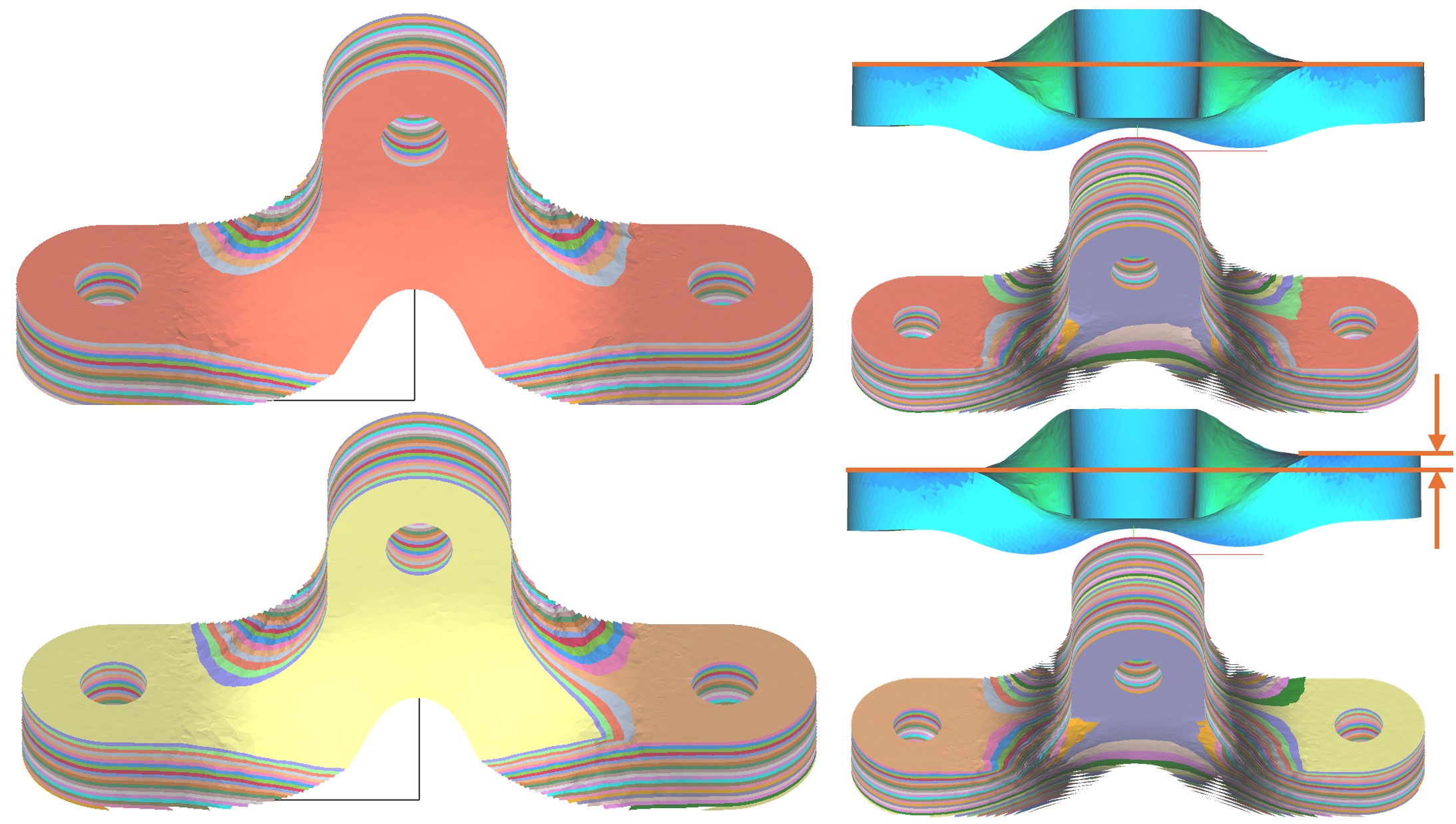}\vspace{-5pt}
\put(-255,135){\footnotesize \color{black}(a)}
\put(-255,60){\footnotesize \color{black}(b)}
\caption{The computational result of T-bracket model: (a) with or (b) without height consistency control. When the height consistency is not controlled, it will not be able to form a complete layer to connect the three holes under loading -- therefore cannot form hole winding toolpaths to strengthen the reinforcement of continuous fibers.
}\label{fig:yConsistency}
\vspace{-5pt}
\end{figure}

Secondly, since the relationships between the holes are not formulated into the optimization problem, the load-bearing layers may be divided into separate regions. This prevents the formation of a continuous surface layer with continuous fibers connecting the holes under loads. To address this issue, we introduce an energy term in the global blending step (refer to Eq.~\eqref{eq:deformASAP}) to ensure that the region-of-interests (ROI) are maintained at the same height in the deformation space, thereby resolving the problem -- see the ablation study to demonstrate the effectiveness of this additional term as shown in Fig.~\ref{fig:yConsistency}.
\begin{equation}\label{eq:deformASAP_hole}
\begin{aligned}
\arg \min_{\mathcal{M}^d} & \ \ E_{pc}(\mathcal{M}^d) + E_{rs}(\mathcal{M}^d) + E_{sc}(\mathcal{M}^d) \\ 
& + \underbrace{ w_{hc}\sum_{(v_i,v_j) \in ROI} ({y}_{v_i} - {y}_{v_j} )^2 }_{\mathrm{Height~Consistency}~E_{hc}(\mathcal{M}^d)},
\end{aligned}
\end{equation}
where ${y}_{v_i}$ represents the height of vertex $v_i$ in the deformation space, and $w_{hc}=2.0$ is the weight used to balance this requirement with other objectives. 

\begin{figure}[t]
\centering  
\includegraphics[width=\linewidth]{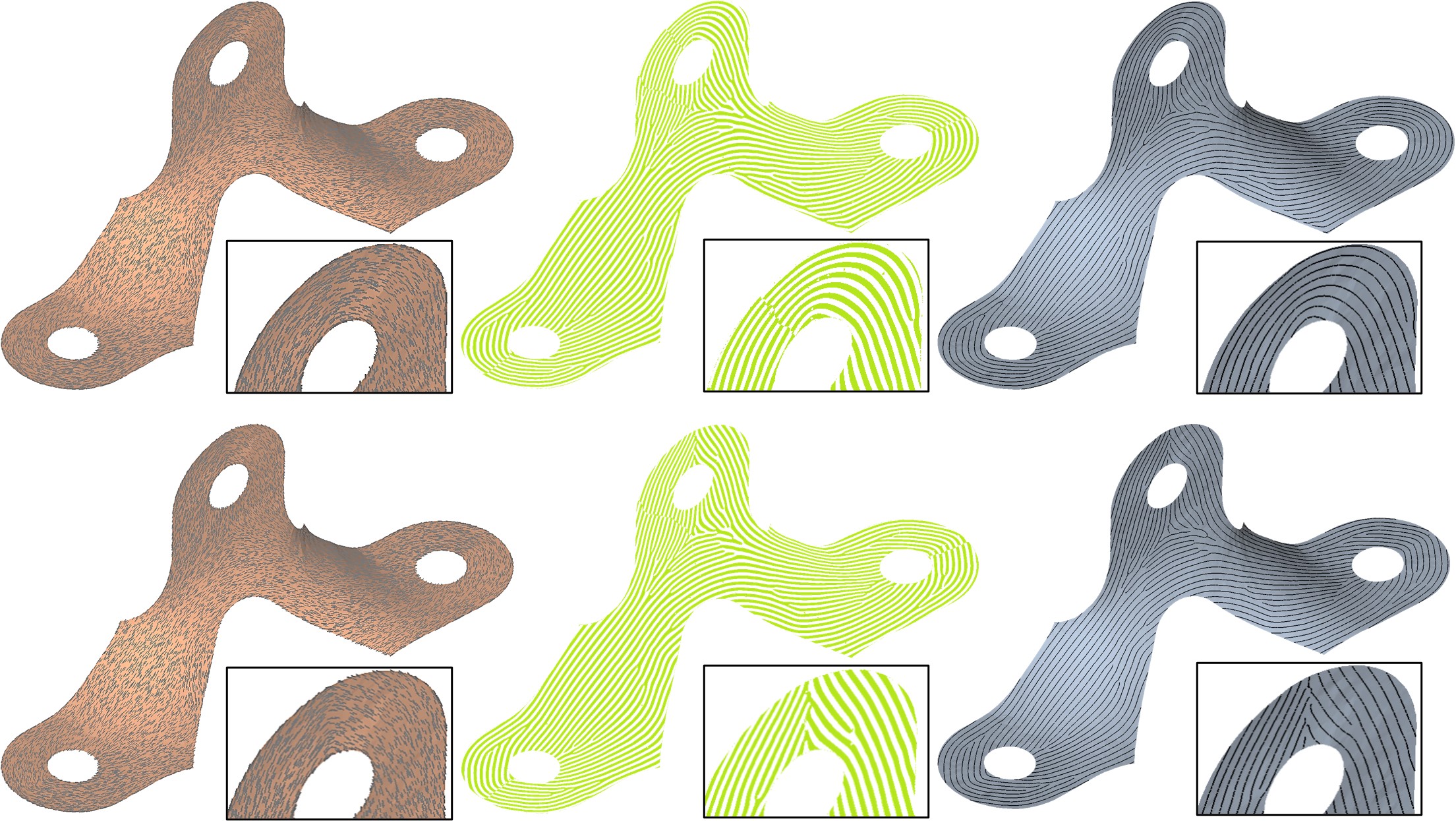}\vspace{-5pt}
\put(-255,135){\footnotesize \color{black}(a1)}
\put(-175,135){\footnotesize \color{black}(a2)}
\put(-90,135){\footnotesize \color{black}(a3)}
\put(-255,60){\footnotesize \color{black}(b1)}
\put(-175,60){\footnotesize \color{black}(b2)}
\put(-90,60){\footnotesize \color{black}(b3)}
\caption{
Comparison to study the effectiveness of the boundary following term $E_{bf}(\mathcal{S})$: In (a1-a3), the direction field incorporates boundary-following, resulting in strips and extracted paths that wrap around the load-bearing holes. Differently, (b1-b3) show the case without considering the boundary following.}
\label{fig:boudaryKept}
\vspace{-5pt}
\end{figure}

Lastly, a boundary following term can be added to further enhance the computation of the 2-RoSy directional field for considering the strip generation around holes. This is realized by enforcing the directions to be consistent with averaged boundary edges' directions on boundary vertices located at the boundary of holes (stored in a set $\mathcal{E}_b$).
\begin{equation}\label{eq:directionFieldOpt_hole}
    \arg \min_{\mathbf{d}_{v_i}}\ \ E_{sa}(\mathcal{S}) + E_{uf}(\mathcal{S}) \\ +
    \underbrace{
     \frac{\omega_h}{|\mathcal{E}_b|}\sum_{e_{ij}\in \mathcal{E}_b} \| \mathbf{d}_{v_i} - \mathbf{b}_{v_i} \|^2_2}_{\text{\rev{}{Boundary~Following}}~E_{bf}(\mathcal{S})},
\end{equation}
where $\mathbf{b}_{v_i}$ represents the boundary directions as a 2-RoSy vector at a boundary vertex $v_i$, and $\omega_h = 55.0$ is a weight in all our examples balancing the multiple objectives. The effectiveness of this boundary following term has been demonstrated by the example shown in Fig.~\ref{fig:boudaryKept}.

\begin{figure*}[t]
\centering
\includegraphics[width=0.99\linewidth]{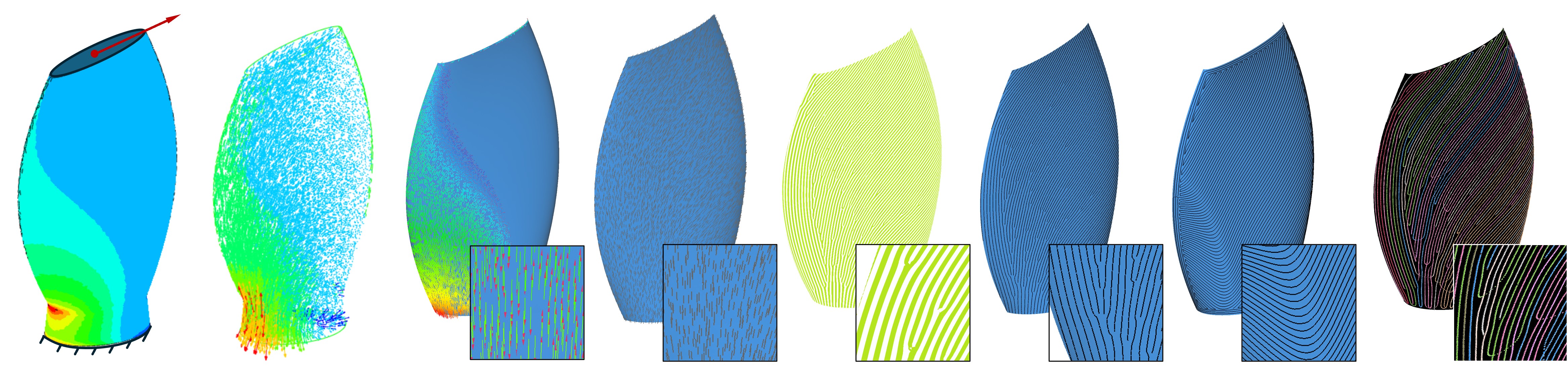}
\put(-530,118){\footnotesize \color{black}(a)}
\put(-500,88){\footnotesize \color{black}Force}
\put(-520,-5){\footnotesize \color{black}Fixed}
\put(-455,118){\footnotesize \color{black}(b)}
\put(-390,118){\footnotesize \color{black}(c)}
\put(-325,118){\footnotesize \color{black}(d)}
\put(-265,118){\footnotesize \color{black}(e)}
\put(-200,118){\footnotesize \color{black}(f1)}
\put(-135,118){\footnotesize \color{black}(f2)}
\put(-70,118){\footnotesize \color{black}(g)}
\caption{Computational results of the blade model: (a) and (b) show the maximum principal stress field of the input model under specified loads as boundary conditions; (c) presents the projection of the stress field onto the curved surface, where an apparent directional conflict can be observed in the zoom-view; (d) shows the optimized directional field $\mathbf{d}$; (e) displays a stripe pattern with equal spacing generated by computing a periodically parameterized scalar field; (f1) depicts the fiber toolpath extracted from the stripe pattern. Compared to (f2), as discussed in~\cite{Fang_SIGA20}, the toolpath generated by our method has equal hatching space and faithfully follows the stress flow direction. Finally, (g) illustrates the toolpath used for manufacturing, obtained by post-processing the toolpath shown in (f1) according to the manufacturing constraints of continuous fibers.
} \label{fig:bladeRes}
\end{figure*}

\section{Results and Discussion}\label{secResults}
This section details the computational results of the proposed fiber toolpath generation method, followed by the fabrication of mechanical parts with varied geometries and stress distributions using an 8-DoF robotic system. Tensile and 3-point bending tests have been conducted to verify the effectiveness of our method in generating spatial fiber toolpaths with high density, which can significantly enhance the reinforcement of mechanical strength. Results of the toolpaths, the fabrication process, and the tensile tests can also be found in the supplementary video at: \url{https://youtu.be/ylBgGtqyhDE}. 

\subsection{Computational Results}
All computational tests were carried out on a PC equipped with an AMD Ryzen 9 7900X CPU (12 cores @ 4.7GHz) and 64GB of RAM, running Windows 11. Sparse matrix computations were accelerated using the Eigen library~\cite{Eigen} and the MKL library~\cite{Wang_ICDM14}. The method's effectiveness has been evaluated on a series of tests using models with freeform geometries subjected to varying load conditions.

The first example conducted in our tests is a T-bracket model as already shown in Fig.~\ref{fig:pipeline}, where the bottom is secured with bolts and the top hole is subjected to an upward force. The axes of the bottom and top holes are arranged vertically. The plane slice cannot obtain the layer surrounding the three holes, and thus cannot fully utilize the reinforcement effect of the continuous fiber. The extended version of $S^3$-slicer can generate curved layers similar to~\cite{Fang_ADDMA24} but does not require the tedious stress line tracing step. The computational pipeline has been shown in Fig.~\ref{fig:pipeline}(a)-(d). The resultant set of curved layers, along with the stress field on the layers, is then used as input for the toolpath generation algorithm, which computes the smooth direction field that aligns with the maximum principal stress (Fig.~\ref{fig:pipeline}(e)). Considering the manufacturing requirement of equidistant fiber placement, stripes with nearly equal hatching distances can be generated by our algorithm (Fig.~\ref{fig:pipeline}(f)). The toolpath is extracted from the stripes (Fig.~\ref{fig:pipeline}(g)) and the final printing toolpath was obtained as shown in Fig.~\ref{fig:pipeline}(h) after post-processing. The spatial fiber toolpath for multi-axis additive manufacturing generated by our computational pipeline can significantly increase the \rev{density}{volume fraction} of fibers while accurately aligning the fiber's toolpaths with the stress directions. This result maximizes the mechanical reinforcement by leveraging the strong anisotropy of the continuous fibers.

\begin{figure}[t]
\centering
\includegraphics[width=\linewidth]{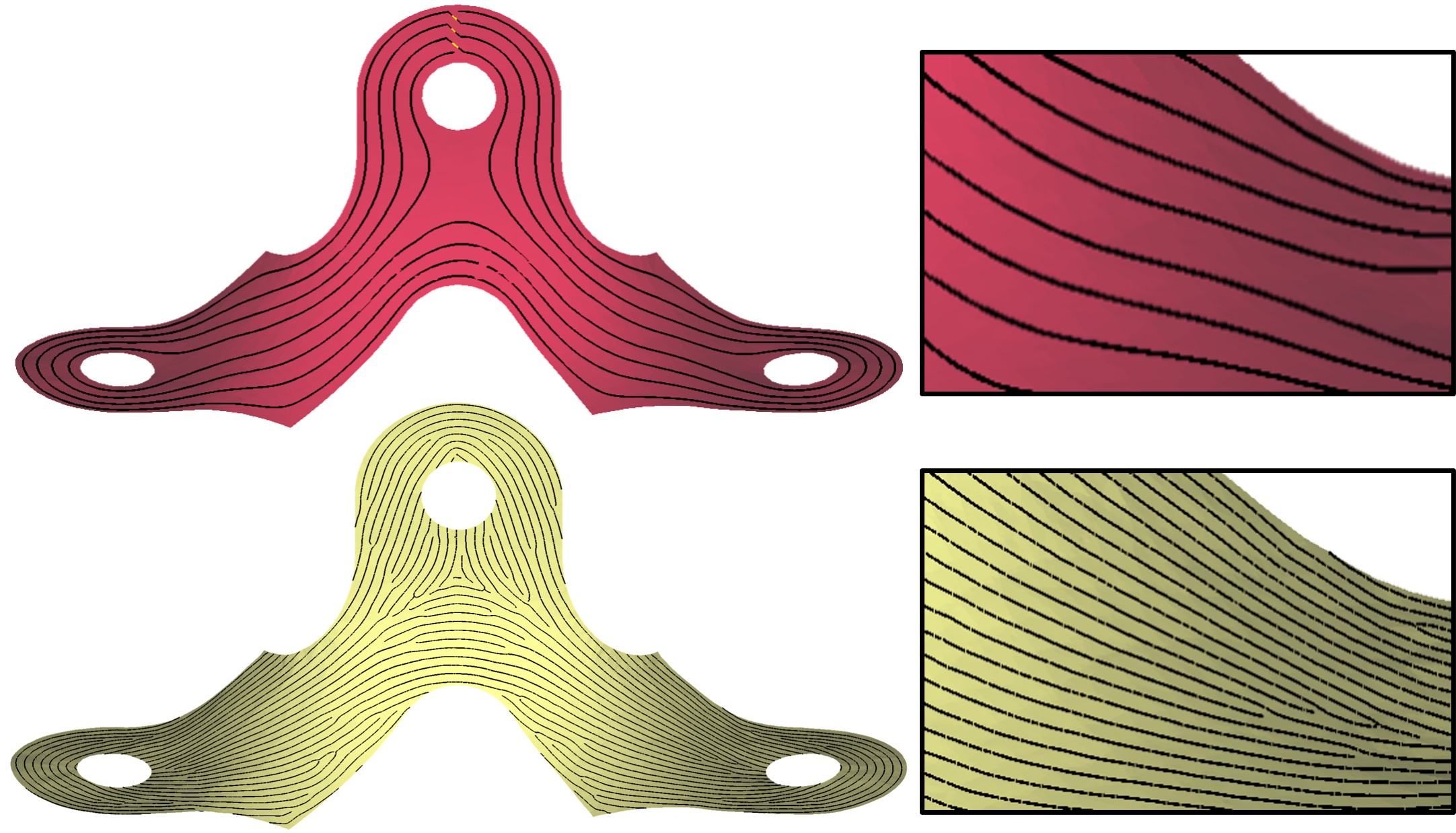}
\put(-255,130){\footnotesize \color{black}(a)}
\put(-255,60){\footnotesize \color{black}(b)}
\caption{Comparison of the results for the T-bracket model: (a) the toolpath generated by our method, and (b) the toolpath from the previous work~\cite{Fang_ADDMA24} -- both are generated on the same curved layer.
}
\label{fig:densityCompare}
\end{figure}

\begin{figure}[t]
\centering
\includegraphics[width=0.99\linewidth]{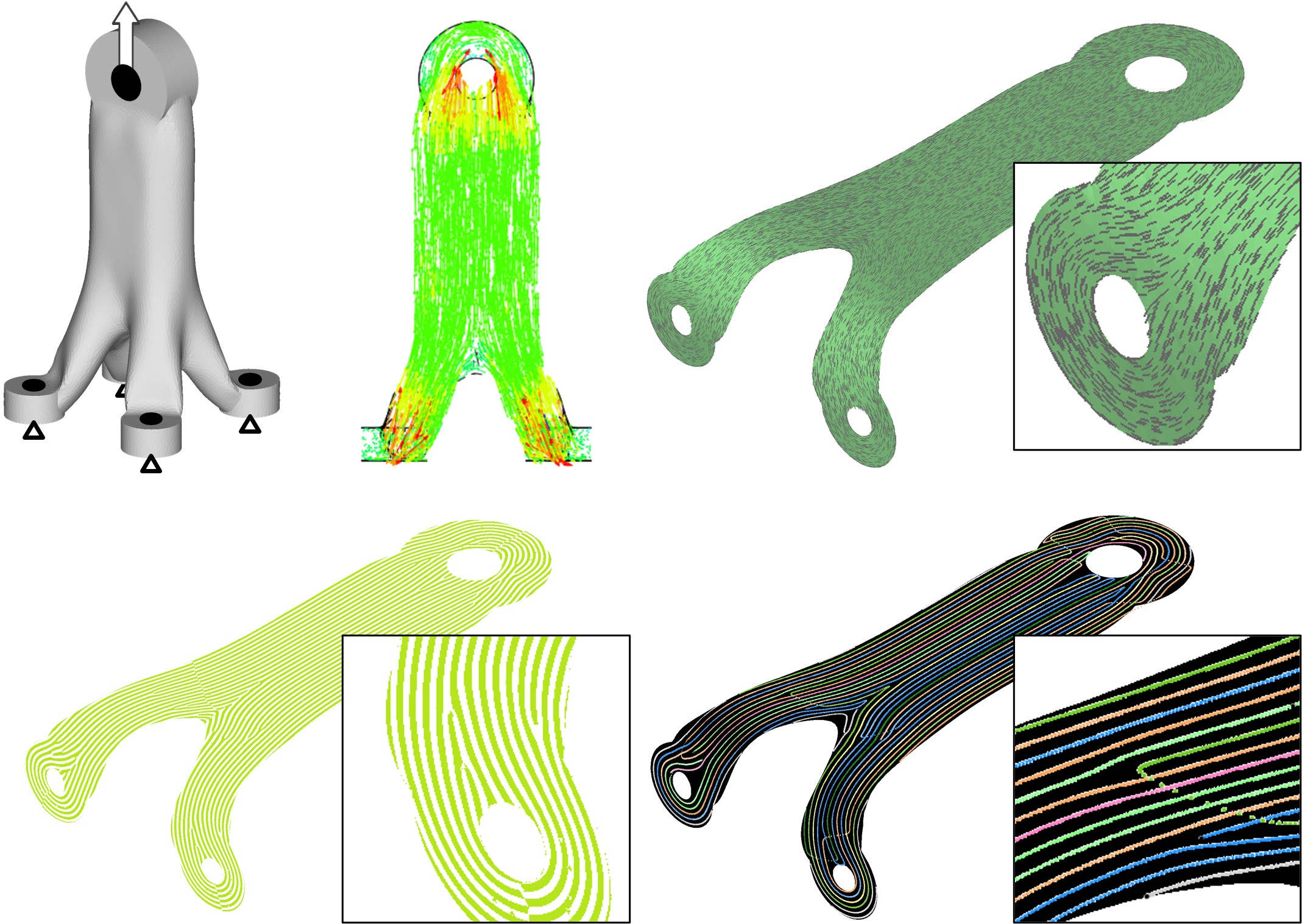}
\put(-259,78){\footnotesize \color{black}(a)}
\put(-213,155){\footnotesize \color{black}Force}
\put(-209,115){\footnotesize \color{black}Fixed}
\put(-195,78){\footnotesize \color{black}(b)}
\put(-130,78){\footnotesize \color{black}(c)}
\put(-259,3){\footnotesize \color{black}(d)}
\put(-130,3){\footnotesize \color{black}(e)}
\caption{The result of X-bracket model: (a) \& (b) the input model and the maximal principle stress field under the given boundary condition; (c) the optimized direction field $\mathbf{d}(\cdot)$ following the stress field; (d) the strip pattern with nearly equal hatching distance; (e) the fiber toolpath extracted from the strip pattern, which has been post-processed by considering the manufacturing constraints.} \label{fig:X_bracketRes}
\end{figure}

\begin{figure}[t]
\centering
\includegraphics[width=0.99\linewidth]{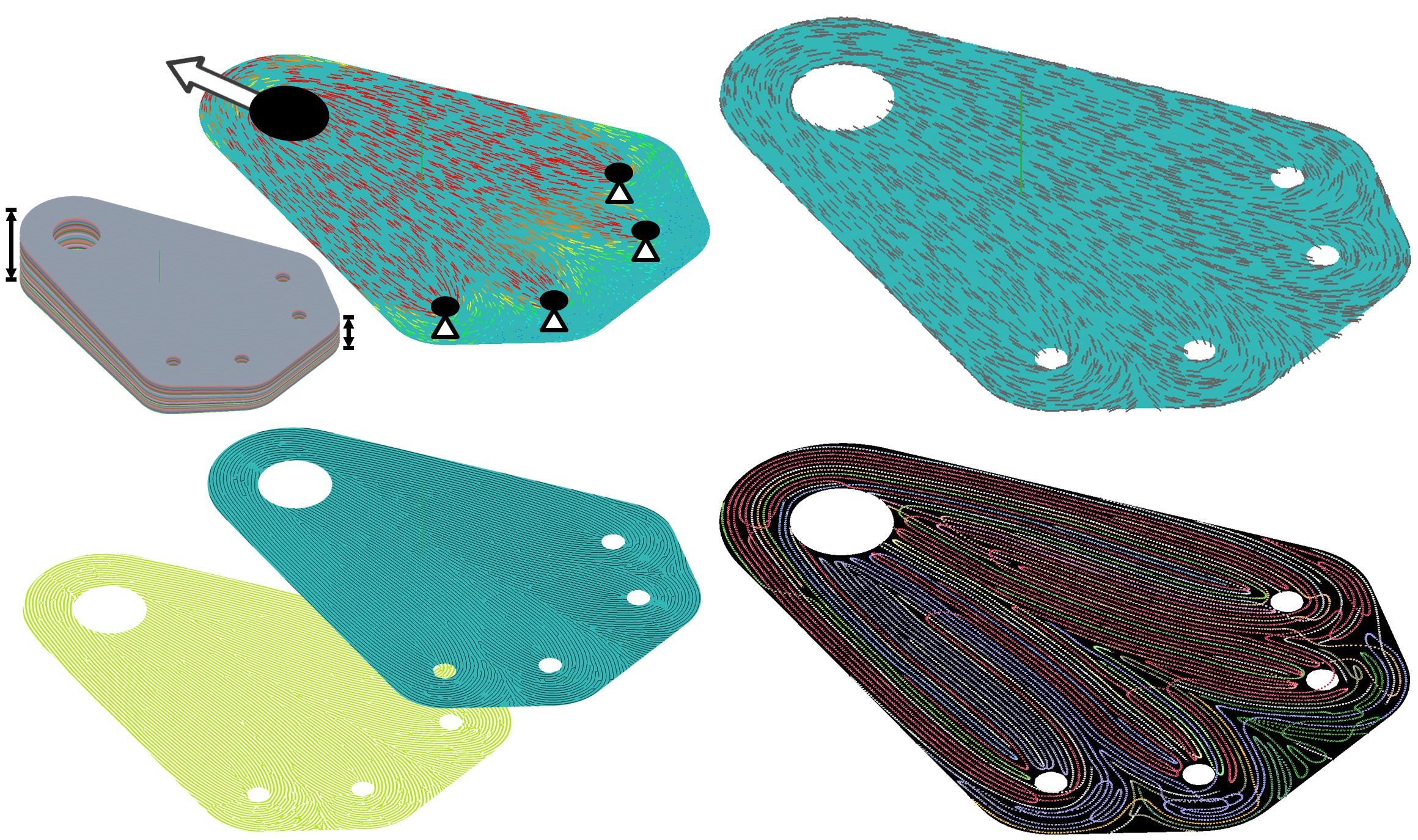}
\put(-259,75){\footnotesize \color{black}(a)}
\put(-246,130){\footnotesize \color{black}Force}
\put(-159,135){\footnotesize \color{black}Fixed}
\put(-130,75){\footnotesize \color{black}(b)}
\put(-259,5){\footnotesize \color{black}(c)}
\put(-130,5){\footnotesize \color{black}(d)}
\caption{The computational results of the Wedge model: (a) the input model and the projected maximal stress field under the given boundary conditions (note that the layer height varies on the model as indicated by the black arrows); (b) an optimized direction field $\mathbf{d}(\cdot)$ that follows the stress field; (c) the strip pattern and the unprocessed toolpath as the iso-curves extracted from the strips; (d) the final fiber toolpath processed by considering the fabrication constraints.} \label{fig:wedgeRes}
\end{figure}

By computing the length of the fiber toolpath and experimentally measuring the fiber's width (1.1~mm in our case), we can evaluate the coverage area of the fiber layer. As shown in Fig.~\ref{fig:densityCompare}, the coverage of the corresponding fiber path is 87.5\%, which is significantly higher than the $26.0\%$ coverage that can be generated by the prior work presented in~\cite{Fang_ADDMA24}. Similar optimization results are observed in the other four components tested in our study, including the Bridge model (Fig.~\ref{fig:bridgeRes}), the Blade model (Fig.~\ref{fig:bladeRes}), the X-bracket model (Fig.~\ref{fig:X_bracketRes}) and the Wedge model (Fig.~\ref{fig:wedgeRes}). Additionally, in the tests that measure the level of alignment between the fiber directions and the stress directions, the average alignment angle was found to be less than $1.16^\circ$, indicating that the calculated continuous fiber toolpath closely follows the stress direction. 

\begin{figure}[t]
\centering  
\includegraphics[width=\linewidth]{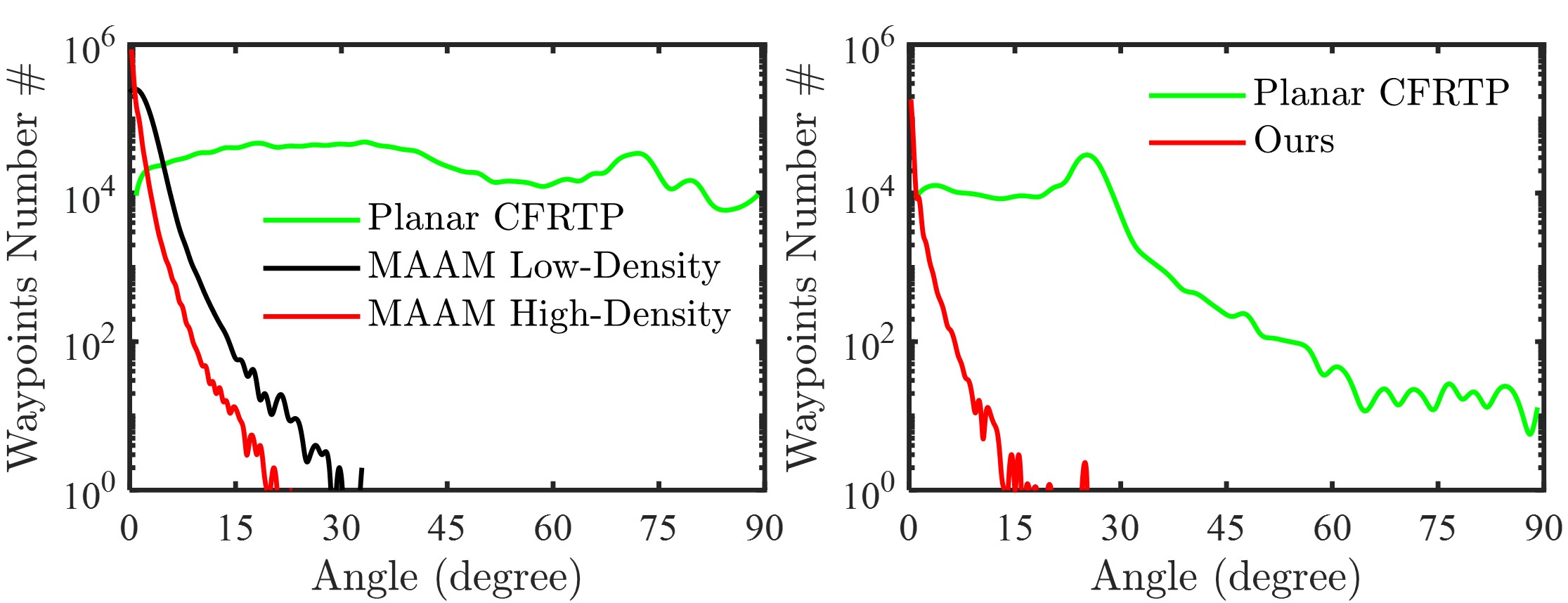}\vspace{-5pt}
\put(-256,0){\footnotesize \color{black}(a)}
\put(-130,0){\footnotesize \color{black}(b)}
\caption{Statistics evaluating the alignment between the tangent direction of the continuous fiber path and the principal stress direction $\sigma_{max}$. (a) For the T-bracket model, compared to our previous method~\cite{Fang_ADDMA24}, the new method can significantly narrow down the range of difference in angle, reducing the maximum from 30.1$^\circ$ to 23.96$^\circ$ and the mean value from 2.54$^\circ$ to 1.16$^\circ$. This indicates the significant improvement of fiber alignment with the principal stress directions. (b) For the Bridge model, our method can limit the angle to $\leq 25.5^\circ$, whereas the planar method yields a wider distribution and therefore fails to fully exploit the fiber's potential for reinforcement.}
\label{fig:histogram_compare}
\vspace{-5pt}
\end{figure}

Our computational pipeline demonstrates high efficiency, as can be observed from the computational statistics listed in Table~\ref{tab:CompRes}. \rev{}{We conducted the computational experiments three times and recorded the average values in this table.} The computation on all examples tested in this paper can be completed within 8 minutes. \rev{}{Note that the resolution of the tetrahedral mesh employed in our method needs to be fine enough to capture the geometric details of an input model. According to the analysis conducted in \cite{Zhang_SIGA22}, further refinement of the tetrahedral mesh will not generate significant influence on the results. The surface meshes employed to generate toolpath are usually remeshed into a resolution comparable to the tetrahedral mesh.}

\begin{table*}[t]
\centering 
\caption{Statistics of generating curved layers and spatial toolpaths.}\label{tab:CompRes}
\footnotesize \vspace{5pt}
\begin{tabular}{r|c|r||c|c||c|c|c|c|c||r}
\hline 
&   &   &  \multicolumn{2}{c||}{Comp. Time of Layers (sec.)}  & Layer \# &  \multicolumn{4}{c||}{Comp. Time of Toolpaths (sec.)$^\dag$}   &   Total Time\\ 
\cline{4-5}  \cline{7-10}
{Model} & {Fig.} &  Tet~\#  & {FEA Comp.} & {$S^3$-slicing} &  Matrix / Fiber & $\sigma_{max}$ Proj. &  $\mathcal{D}(\textbf{x})$ Comp.  & Strip Gnen. & Toolpath Extr. &  {(sec.)} \\ 
\hline\hline \specialrule{0em}{1pt}{1pt}
T-bracket &  ~\ref{fig:pipeline} &  85,119 & 24.23 & 62.46 &  70 / 20 &  12.49 &  28.75 & 43.58 & 62.82 & 234.33
\\ \specialrule{0em}{1pt}{1pt}
Bridge & ~\ref{fig:bridgeRes} & 197,395 & 52.97 & 115.38 & 80 / 20 & 59.11 & 15.12 & 24.38 & 183.63 & 450.59 \\ \specialrule{0em}{1pt}{1pt}
Blade & ~\ref{fig:bladeRes} & 211,979 & 56.77 & / & 4 / 2 & 3.57 & 2.74 & 6.08 & 24.60 & 93.76 \\ \specialrule{0em}{1pt}{1pt} 
X-bracket &  ~\ref{fig:X_bracketRes} & 315,904 & 83.52 & 109.36 & 66 / 12 & 28.52 & 27.88 & 26.38 & 97.20 & 372.86 \\ \specialrule{0em}{1pt}{1pt}
Wedge &  ~\ref{fig:wedgeRes} & 130,282 & 33.83 & 57.59 & 30 / 30 & 73.05 & 6.65 & 11.81 & 29.67 & 212.60 \\\hline
\end{tabular}
\vspace{-5pt}
\begin{flushleft}
~~~~$^\dag$~This includes computing time for generating toolpaths for both the matrix material and the continuous fibers on all curved layers.
\end{flushleft}
\vspace{-10pt}
\end{table*}

\subsection{FEA-Based Validation}
\begin{figure}[t]
\centering
\includegraphics[width=\linewidth]{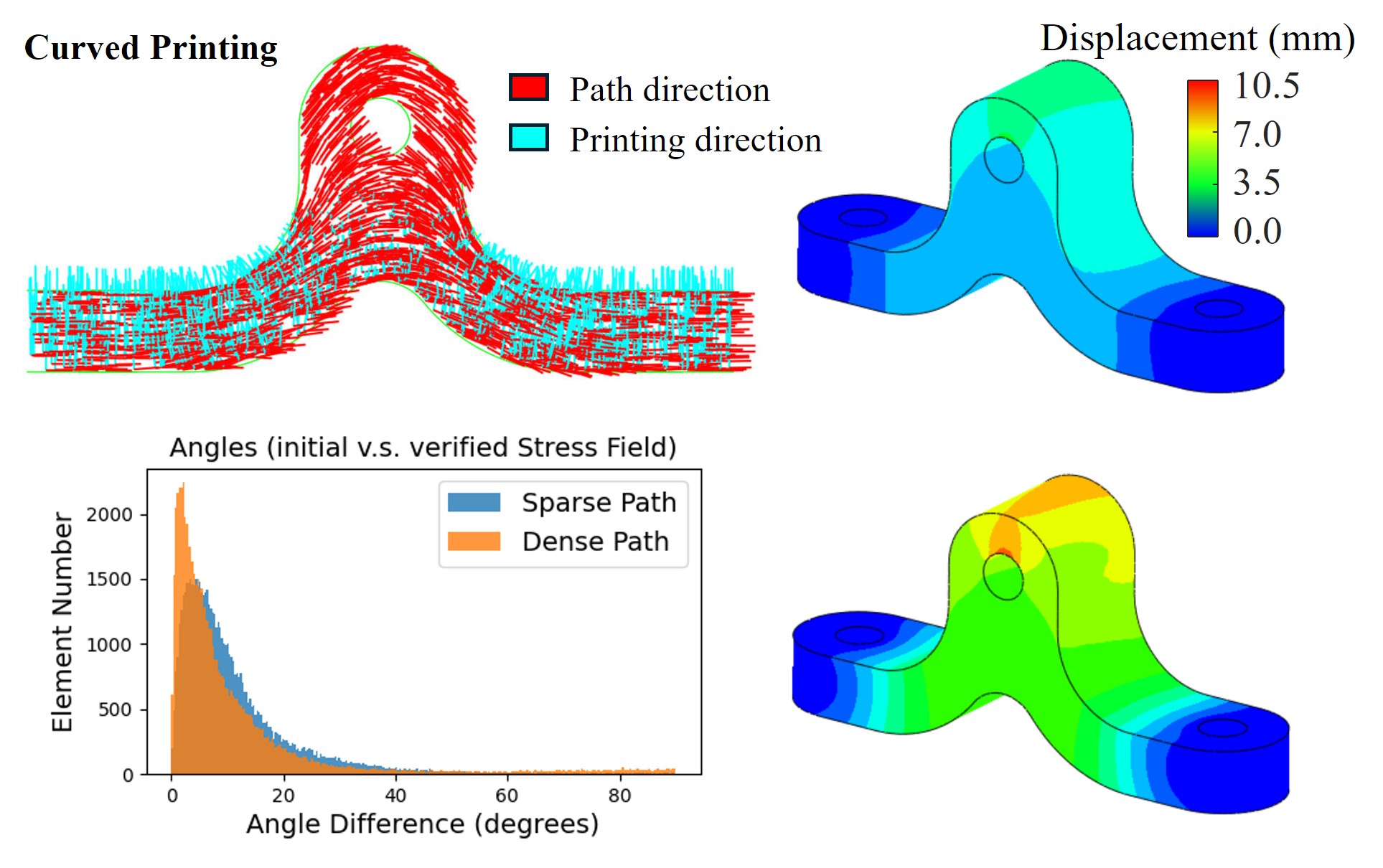}
\put(-256,80){\footnotesize \color{black}(a)}
\put(-120,80){\footnotesize \color{black}(b) $u_{\max}=4.28\text{mm}$}
\put(-256,-2){\footnotesize \color{black}(c)}
\put(-120,-2){\footnotesize \color{black}(d) $u_{\max}=8.94\text{mm}$}
\caption{
\rev{}{Anisotropic FEA is conducted to verify the effectiveness of fiber paths generated by our method with directions shown as red arrows in (a) w.r.t. the local printing directions shown as light blue arrows. The FEA result with dense fiber paths is shown as (b) the distribution of displacement magnitude, and the corresponding FEA result with sparse fiber paths is given in (d). For the case with sparse fiber paths, the anisotropic material property is only applied to the elements containing fiber paths. The maximal displacement of the composite using sparse paths is $108.9\%$ larger than ours by dense paths. The distributions of the angles between the principal stresses obtained from the anisotropic FEA and the fiber path are computed and shown in (c). 
}
}\label{fig:FEA_verfication}
\end{figure}

\rev{}{Before conducting the physical experiments based validation, we have employed the commercial FEA software, Abaqus, to verify the effectiveness of fiber paths generated by our method. Firstly, the composite with dense fiber paths is compared with the one with sparse fiber paths generated by \cite{Fang_ADDMA24} as shown in Fig.~\ref{fig:FEA_verfication}. The anisotropic material orientations are assigned by the toolpath direction $\mathbf{t}$, the local printing direction $\mathbf{g}$, and the third direction as $\mathbf{t} \times \mathbf{g}$. The anisotropic material properties are assigned as Young's moduli $(E_1,E_2,E_3)=(200.0\text{GPa},10.0\text{GPa},10.0\text{GPa})$ and the Poisson's ratio $\mu=0.30$ for those elements containing CCF paths. For the case with sparse fiber path, there are elements for matrix materials without CCF, which are assigned with $E_1=E_2=E_3= 2.7\text{GPa}$ and $\mu=0.33$ as PLA. The anisotropic FEA results prove that the maximal displacement in the composite by sparse fiber paths as $8.94\text{mm}$ is $108.9\%$ larger than ours by dense paths as $4.28\text{mm}$.}

\rev{}{An interesting study is to check the angles between the fiber paths and the maximum principal stresses. The principal stresses given as input of our toolpath generation method are calculated by isotropic FEA, where anisotropic FEA with given CCF directions will theoretically generate principal stresses in changed orientation. From the histograms of the angles as shown in Fig.~\ref{fig:FEA_verfication}(c), we can observe that the angles are small in general with the average as $11.06^{\circ}$ for dense paths, which is slightly lower than that obtained from the sparse path as $11.76^{\circ}$. The strategy of using principal stresses obtained from isotropic FEA has also been employed in other research approaches~\cite{Kipping2024AM, Fang_ADDMA24, Zhang_SIGA22, Fang_SIGA20}.}

\begin{figure}[t]
\centering
\includegraphics[width=\linewidth]{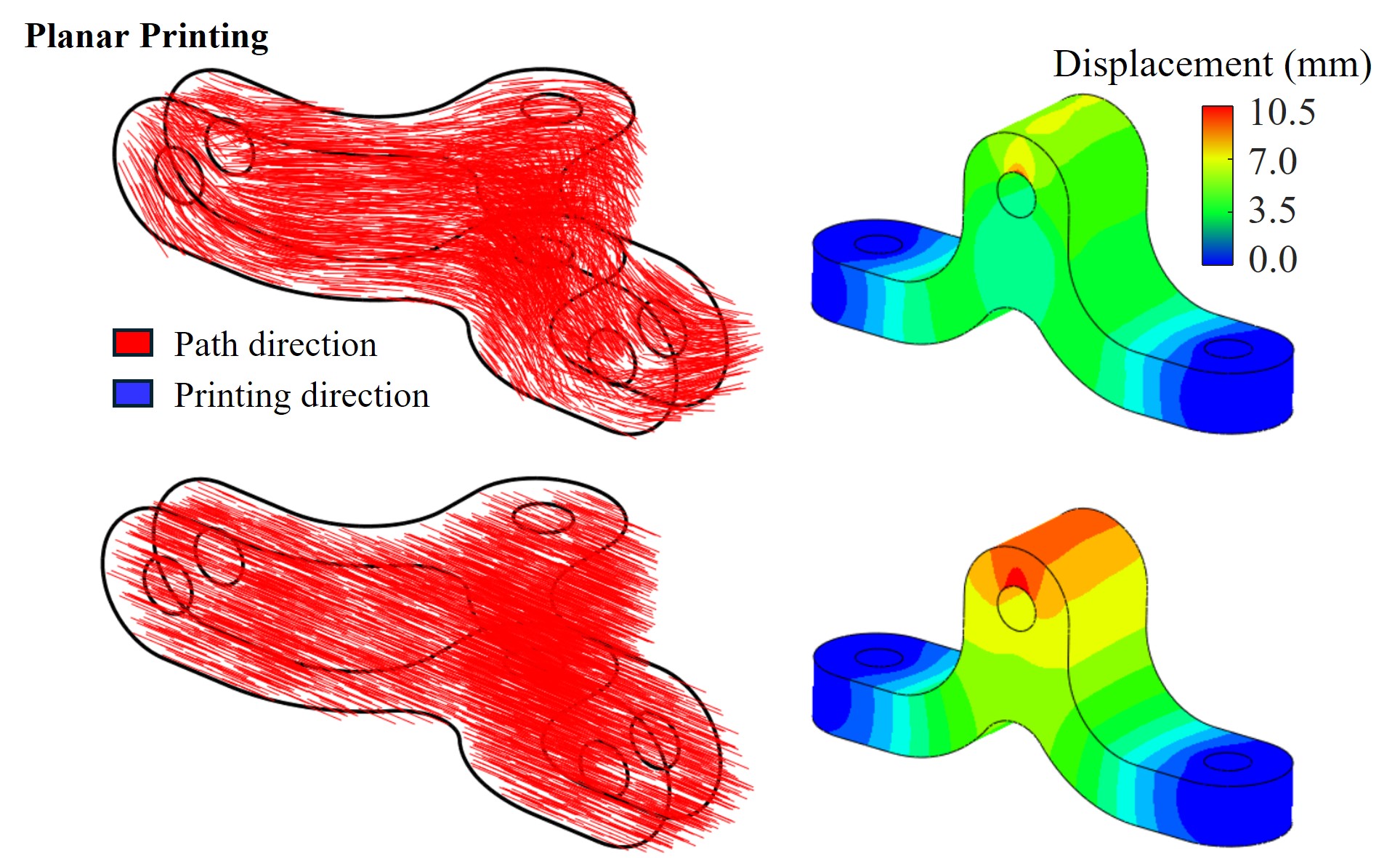}
\put(-256,75){\footnotesize \color{black}(a)}
\put(-125,75){\footnotesize \color{black}(b) $u_{\max}=9.27\text{mm}$}
\put(-256,0){\footnotesize \color{black}(c)}
\put(-125,0){\footnotesize \color{black}(d) $u_{\max}=10.42\text{mm}$}
\caption{
\rev{}{Anisotropic FEA has also been conducted for composites made by planar layers using (a) the stress-field-based planar paths generated by our method and (c) the simple zig-zag paths. The FEA results as by the displacement magnitude as (b) for the stress-field-based planar paths and (d) for the zig-zag planar paths.}
}\label{fig:FEA_verfication2}
\end{figure}

\rev{}{We also conduct the anisotropic FEA to compute the displacement for composites with planar layers of CCF but different patterns of fiber paths. The results are given in Fig.~\ref{fig:FEA_verfication2}. It can be observed that our method can also improve the mechanical strength even when planar based printing is employed. The paths generated by our method can reduce the displacement by $11.2\%$ compared to the standard zig-zag toolpaths used in industry. When comparing with our method by curved layers with dense fiber path, the displacement has been reduced by $58.9\%$.}


\subsection{Hardware and Manufacturing Parameters}\label{subsecHardware}
\begin{figure}[t]
\centering
\includegraphics[width=\linewidth]{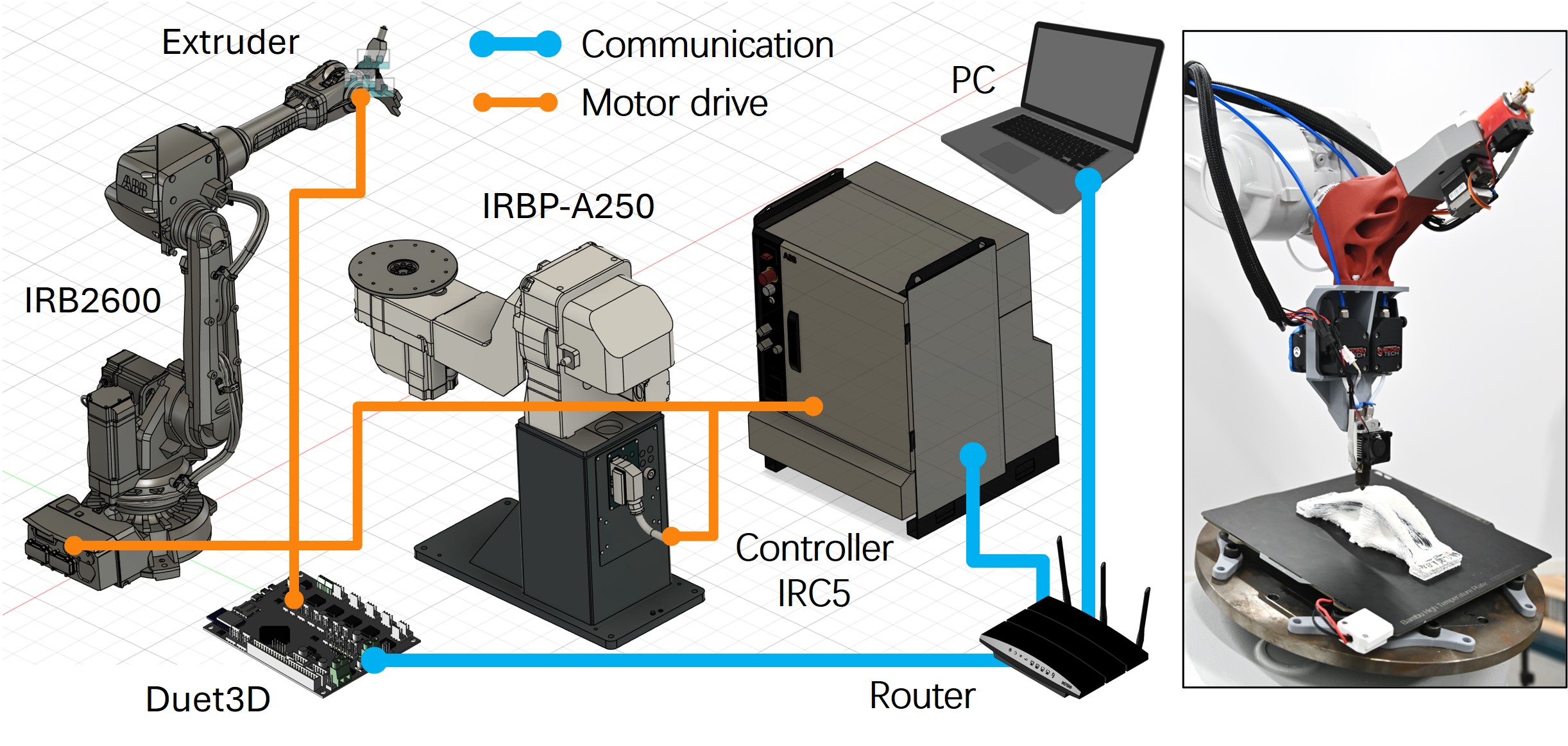}
\put(-253,0){\footnotesize \color{black}(a)}
\put(-73,0){\footnotesize \color{black}(b)}
\caption{Illustration of our robotic hardware setup for continuous fiber fabrication. (a) The 8-DoF robot-assisted additive manufacturing system, where the router functions as an information hub that enables the simultaneous control of both the robots and the extrusion processes. (b) Our system has two printer heads with one for continuous fibers and the other for matrix and supporting materials.} 
\label{fig:hardware}
\end{figure}

An 8-DoF robotic additive manufacturing setup is employed to realize the spatial alignment of continuous fibers. The fabrication process utilizes an out-of-nozzle impregnation strategy, where the matrix material and the continuous fibers are aligned sequentially. To enable multi-axis motion, a 6-DoF ABB 2600 robot arm with a repeatability precision of 0.04mm is used together with a 2-DoF A250 positioner featuring a repeatability precision of 0.05$^\circ$. The robot arm is equipped with a dual-extruder printer head, developed in our previous work~\cite{Zhang_ICRA23}. The printer head allows the extrusion of materials for both the model (e.g., PLA) and the supporting structures (e.g., polyvinyl alcohol). Additionally, a continuous fiber printer head has been installed on the 6-DoF robotic arm. The printer heads are switched alternately to print different materials during fabrication processing (as shown in Fig.~\ref{fig:hardware}(b)). \rev{}{During multi-axis 3D printing, the orientation of the printer head is carefully controlled to align with the surface normal of the substrate.} For all examples shown in this paper, curved layers for the supporting structures were generated by the method proposed in~\cite{Zhang_ICRA23}.

For the fabrication of the models, Bambu 1.75 mm PLA is used for the matrix material, and Markforged CF-FR-50 (with a cross-sectional diameter of 0.37 mm) is selected for the fibers. A nozzle with a diameter of 0.8 mm is employed for extruding the matrix material, with a layer height of approximately 0.6 mm, and the printing temperature is maintained at 220$^{\circ}$C. The continuous fiber printing head is equipped with a nozzle with rounded corners to enable the function of squeezing the fiber filament into the resin matrix already printed, with a layer height of 0.12 mm and a printing temperature of 230$^{\circ}$C.

The robot-assisted fabrication system is controlled by an ABB IRC5, which synchronizes the robot's motion with material extrusion. The extrusion process is managed by a Duet3D board running RepRapFirmware 3.x. The toolpath, represented by a series of waypoints with both position and orientation data, is used to compute the robot joint angles via the method described in~\cite{Zhang_RAL21}, and these are then translated into the RAPID language used in the ABB's RoboStudio software. During the fabrication process, the end-effector's tangential speed is consistently set at around 30 mm/min for thermoplastic material (with adjustment to control the volume of material deposition to be adaptive to the layer thickness variation) and 5 mm/min for continuous fiber material.

\subsection{Results of Fabrication and Mechanical Tests}

\begin{figure}[t]
\centering
\includegraphics[width=\linewidth]{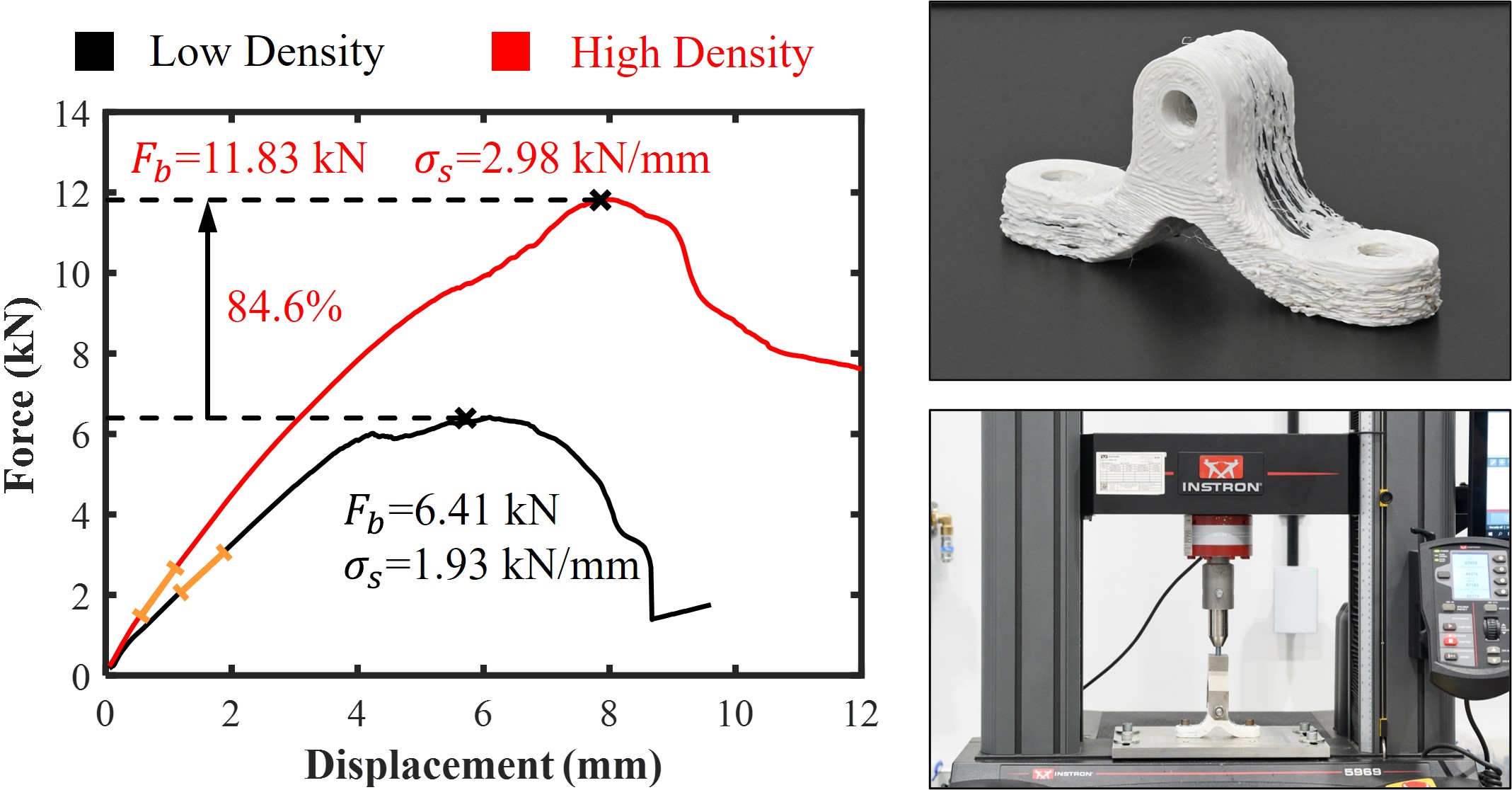}
\put(-256,2){\footnotesize \color{black}(a)}
\put(-115,2){\footnotesize \color{black}(b)}
\caption{Tensile test curves and hardware settings using our toolpaths vs. the low-density toolpaths generated by the prior work~\cite{Fang_ADDMA24}. It can be observed that the fracture force ($F_b$) and the model stiffness are improved by $84.6\%$ and $54.4\%$ respectively.
}
\label{fig:tensileTest_Tbracket}
\end{figure}

The manufacturing results using the high density fiber toolpaths proposed in this paper and the toolpaths generated by our prior work~\cite{Fang_ADDMA24} are compared via mechanical tests. The results are as shown in Fig.~\ref{fig:tensileTest_Tbracket}. The model's size is $130.0\mathrm{mm} \times 30.0\mathrm{mm} \times 60.0\mathrm{mm}$, the matrix materials employed in both specimens are the same as $95.7\mathrm{g}$, and the lengths of the continuous carbon fibers used are $65.9\mathrm{m}$ and $33.7\mathrm{m}$ respectively. It can be found that the fiber density obtained by using the new toolpath is much higher. The direct advantage of this is the significantly improved mechanical strength and stiffness. To verify the mechanical properties of the 3D printed parts, an Instron 5969 dual-column test system with a working range of $30\mathrm{kN}$ was used to conduct the tensile tests. The two main measurements are the fracture force ($F_b$) and the mechanical stiffness (computed as the ratio between force and displacement, $\sigma_s$). As a result, models manufactured using our new toolpath can achieve an $84.6\%$ larger failure load and a $54.4\%$ improvement in mechanical stiffness.

\begin{figure}[t]
\centering
\includegraphics[width=\linewidth]{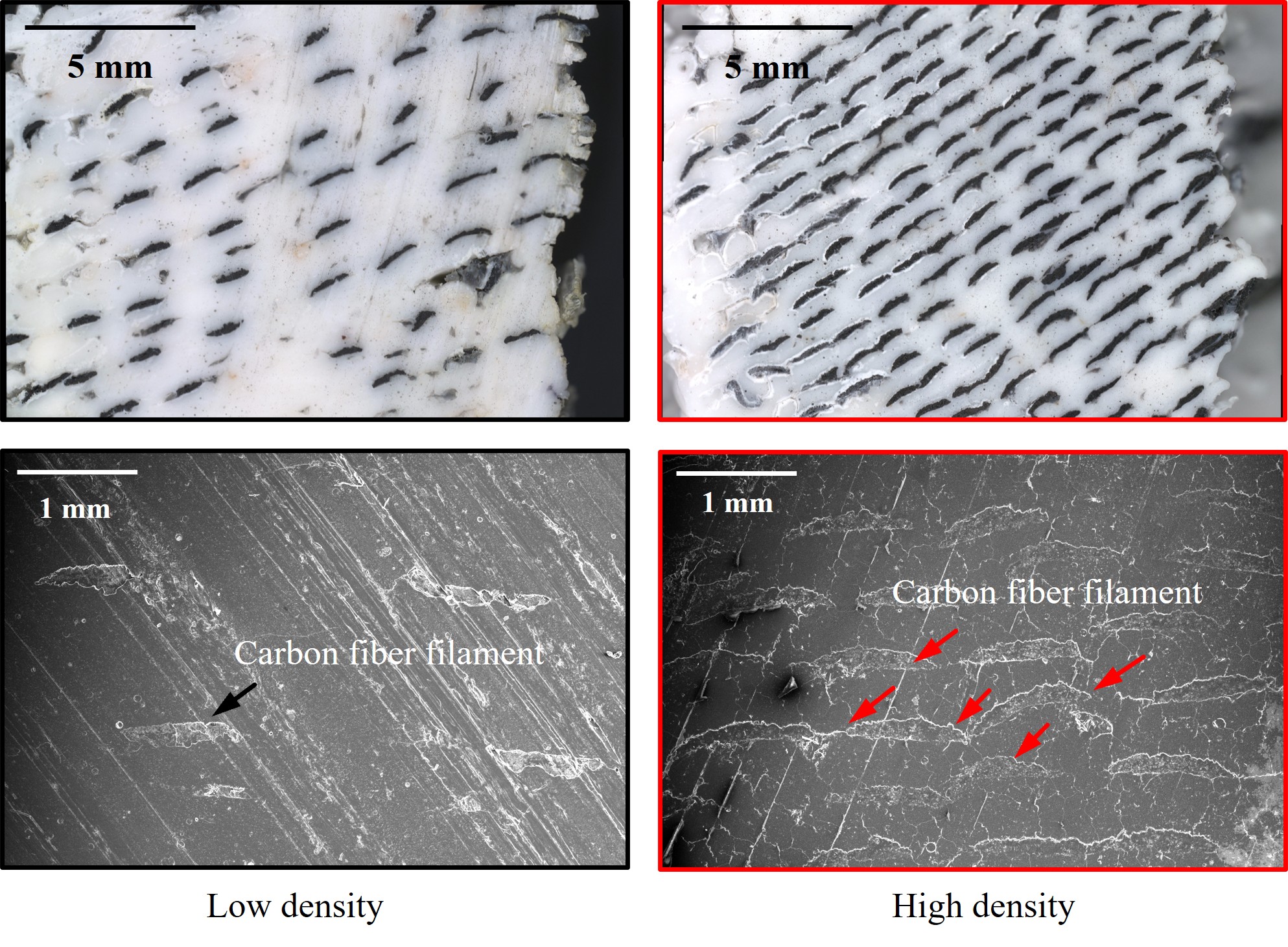}
\put(-255,109){\footnotesize \color{black}(a1)}
\put(-124,109){\footnotesize \color{black}(a2)}
\put(-255,19){\footnotesize \color{white}(b1)}
\put(-124,19){\footnotesize \color{white}(b2)}
\vspace{-5pt}
\caption{The optical microscope and SEM images of cross-sections taken from the low density (a1-a2) vs. the high density (b1-b2) carbon fiber composites. The continuous fiber-reinforced thermoplastic composite built from toolpaths generated by our method shows a much denser distribution of fibers.
}\label{fig:SEM_Tbracket}
\end{figure}

We further validated our results using the optical microscope and scanning electron microscopy (SEM) on the T-bracket model as shown in Fig.~\ref{fig:tensileTest_Tbracket}. The cross-sectional images of low density carbon fiber composites fabricated by the prior method~\cite{Fang_ADDMA24} are given in Fig.~\ref{fig:SEM_Tbracket}(a1 \& b1). The corresponding images of the high density carbon fiber composites generated by our new method are given in Fig.~\ref{fig:SEM_Tbracket}(a2 \& b2). It can be clearly observed that the composite fabricated by using our toolpaths has a much higher density of carbon fibers.

%
%

\begin{figure}[t]
\centering
\includegraphics[width=\linewidth]{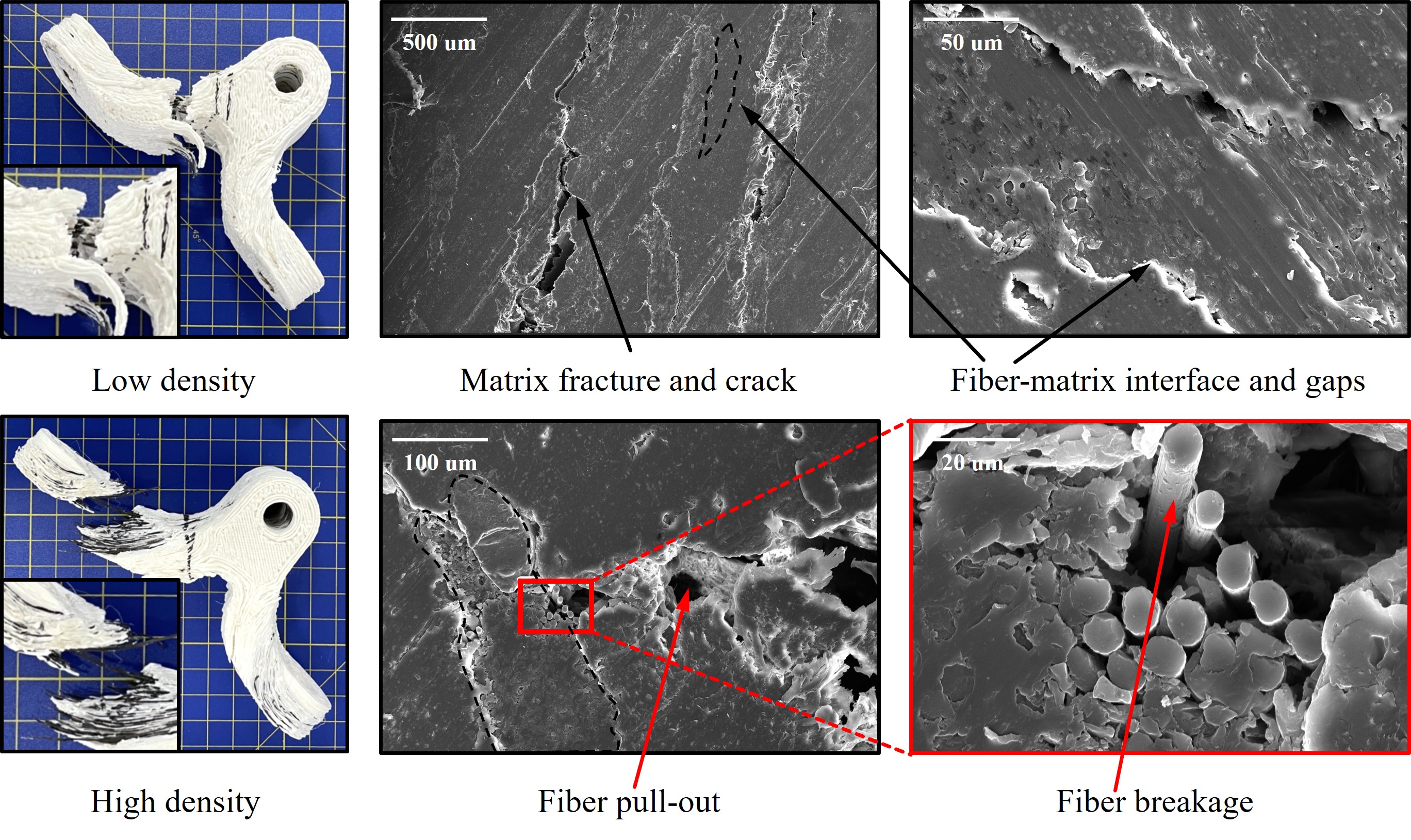}
\put(-256,94){\footnotesize \color{white}(a)}
\put(-256,18){\footnotesize \color{white}(b)}
\put(-187,94){\footnotesize \color{white}(c.1)}
\put(-187,18){\footnotesize \color{white}(c.3)}
\put(-90,94){\footnotesize \color{white}(c.2)}
\put(-90,18){\footnotesize \color{white}(c.4)}
\vspace{-5pt}
\caption{\rev{}{Comparison of failure behavior of (a) the CFRTP composite fabricated from low-density fiber paths (i.e., those generated by \cite{Fang_ADDMA24}) vs. (b) the composite from our high-density fiber paths. The failures in both cases have shown the situations of (c.1) matrix fracture cracks, (c.2) delamination, (c.3) fiber pull-out and (c.4) fiber breakage. However, the failures on CFRTP composite made by dense fiber path have significantly increased occurrence of fiber pull-out.}
}\label{fig:failure_behavior}
\end{figure}

\rev{}{We now examine the failure modes of composites fabricated using our method compared to those produced with low-density fiber paths (e.g., \cite{Fang_ADDMA24}). The comparison has been given in Fig.~\ref{fig:failure_behavior}. As can be observed, both approaches exhibit failure features such as gaps and cracks between the matrix and fibers, fiber pull-out as well as fiber breakage. However, a distinct characteristic of composites made by our method is the increased occurrence of fiber pull-out failures, which are accompanied by delamination (refer to Fig.~\ref{fig:failure_behavior}(c.2 \& c.3)). This is caused by the reduced bonding strength between the fibers and the matrix as discussed in \cite{Caminero_PT18} that higher fiber volume fractions are shown to diminish interfacial bonding performance.}

\begin{figure}[t]
\centering
\includegraphics[width=\linewidth]{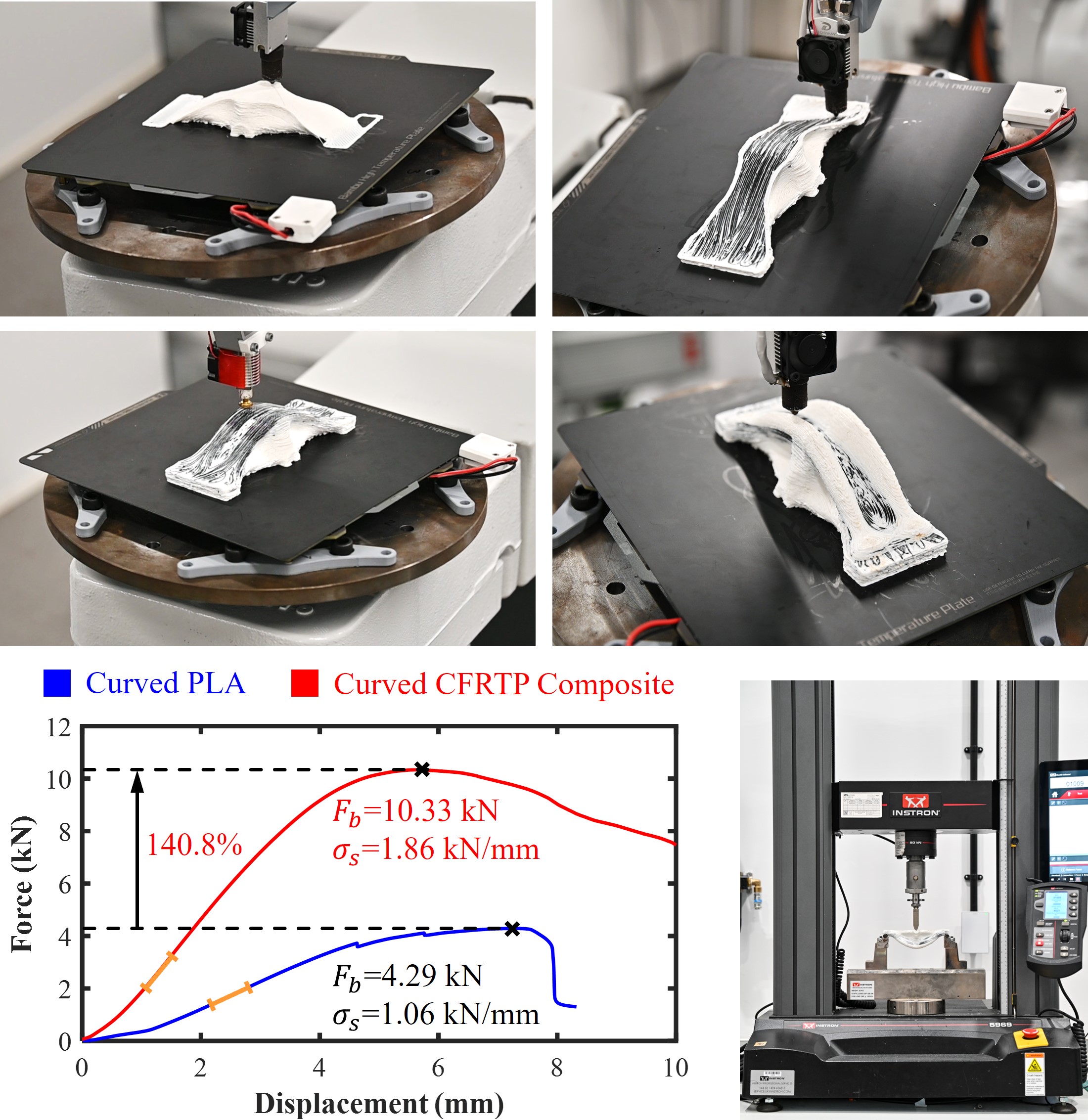}
\put(-255,195){\footnotesize \color{black}(a1)}
\put(-124,195){\footnotesize \color{white}(a2)}
\put(-255,116){\footnotesize \color{black}(a3)}
\put(-124,116){\footnotesize \color{black}(a4)}
\put(-255,-2){\footnotesize \color{black}(b1)}
\put(-100,-2){\footnotesize \color{black}(b2)}
\caption{Illustration of the physical fabrication and test of the Bridge model. (a) Fabrication process using the 8-DoF robotic system. (b) Results and hardware setup of the 3-point bending test comparing the mechanical strength of the CFRTP composite with high density fibers using our toolpath vs. the model fabricated by only using PLA in curved layers. $140.8\%$ improvement can be observed in the failure load and $84.9\%$ increase in stiffness.
}\label{fig:bridgeFab}
\end{figure}

The manufacturing results of the Bridge model are shown in Fig.~\ref{fig:bridgeFab}(a1-a4), with the dimensions of $170.0\mathrm{mm} \times 60.0\mathrm{mm} \times 48.5\mathrm{mm}$, and  $135.7\mathrm{m}$ fibers and $234.8\mathrm{g}$ matrix materials are used respectively. It can be observed that the continuous fiber is closely fitted with the matrix material. The part generated by using our new toolpath can significantly improve the coverage density of fibers. The effectiveness of our method is further validated by the 3-point bending test shown in Fig.~\ref{fig:bridgeFab}(b). Our method achieves a $140.8\%$ enhancement in failure load and a $84.9\%$ improvement in stiffness.

\begin{figure}[t]
\centering
\includegraphics[width=\linewidth]{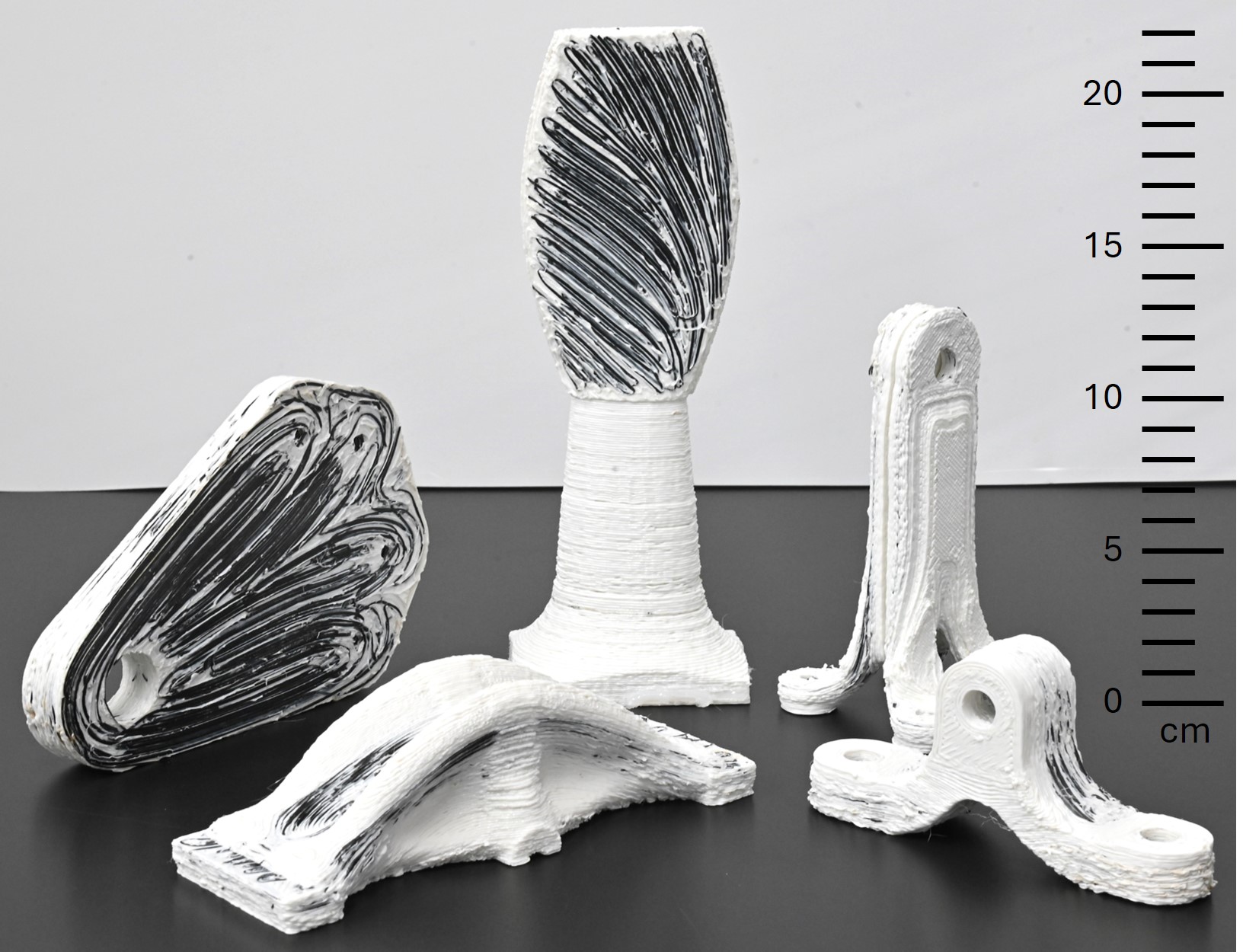}
\put(-250,30){\footnotesize \color{white}(a)}
\put(-165,70){\footnotesize \color{white}(b)}
\put(-95,70){\footnotesize \color{white}(c)}
\put(-230,5){\footnotesize \color{white}(d)}
\put(-95,20){\footnotesize \color{white}(e)}
\caption{All models that have been fabricated by using our toolpath -- (a) Wedge, (b) Blade, (c) X-bracket, (d) Bridge, and (e) T-bracket. 
}\label{fig:resFab}
\end{figure}

Fig.~\ref{fig:resFab} shows all the models we fabricated, with the accompanying table providing the relevant printing parameters and times in Table ~\ref{tab:resFab}. The fabrication time of a model depends on its dimension and also the complexity of the layers and toolpaths. However, the computing time for toolpath generation in general is much shorter than the fabrication time.

\begin{table}
\caption{Fabrication statistics on models fabricated by our method.}
\vspace{5pt}
\centering\label{tab:resFab}
\scriptsize 
\begin{tabular}{r|c|c|c|c|c}
\hline
\specialrule{0em}{1pt}{1pt}
& Dimensions & \multicolumn{2}{c|}{Material Usage} & \rev{}{Fiber Vol.} & Fab. \\
\specialrule{0em}{1pt}{1pt}
\cline{3-4} 
\specialrule{0em}{1pt}{1pt}
Models & ($mm^3$) & Mat. (g) & Fib. (m) & \rev{}{Fr. (\%)} & time (h)\\
\specialrule{0em}{1pt}{1pt} \hline\hline \specialrule{0em}{1pt}{1pt}
T-bracket & $130.0 \times 30.0 \times 60.0 $ &  $95.7$  &  $65.9$ & \rev{}{$8.41$} & $33.5$ \\ 
\specialrule{0em}{1pt}{1pt}
X-bracket & $70.0 \times 70.0 \times 134.1$ &  $119.2$  &  $46.3$ & \rev{}{$4.92$} & $35.7$ \\ 
\specialrule{0em}{1pt}{1pt}
Wedge & $157.5 \times 114.7 \times 29.7$ &  $206.3$  &  $320.1$ & \rev{}{$17.14$} & $15.4$ \\ 
\specialrule{0em}{1pt}{1pt}
Blade\rev{}{$^\ddag$} & $80.0 \times 60.0 \times 220.0$ & $31.7$  &  $24.8$ & \rev{}{$12.2$} & $7.8$ \\
\specialrule{0em}{1pt}{1pt}
Bridge & $170.0 \times 60.0 \times 48.5$ & $234.8$  &  $135.7$ & \rev{}{$7.15$} & $23.7$ \\
\specialrule{0em}{1pt}{1pt}
\hline
\end{tabular}
\begin{flushleft}
\rev{}{$^\ddag$~The Blade model is fabricated by printing 2 layers of CCF + 4 layers of matrix material on top of a base model.}
\end{flushleft}
\vspace{-5pt}
\end{table}

\section{Conclusion}\label{secConclusion}
In this paper, we introduce a novel toolpath generation approach for continuous fiber-reinforced thermoplastic composites made by multi-axis additive manufacturing that is guided by stress fields and optimized for high fiber coverage. Our method overcomes the limitations of existing techniques by incorporating a 2-RoSy direction field to resolve the ambiguity of the principal stress field and using periodic parameterization to generate toolpaths with nearly equal hatching distance. The extension of the $S^3$-Slicer facilitates the generation of curved layers, incorporating winding compatibility considerations.

By aligning fibers along the principal stress directions and optimizing the fiber coverage, our approach can significantly enhance the mechanical strength of CFRTP composites. The results obtained from tensile and 3-point bending tests, combined with SEM analysis, verify the effectiveness of our method in improving both fiber alignment and mechanical strength. Additionally, the implementation of our approach on an 8-DoF robotic system highlights its practical applicability in advanced manufacturing systems, enabling the fabrication of complex, high-performance composite parts.

However, certain limitations still remain in the current approach. \rev{}{Due to the challenge of printing process, the volume fraction of carbon fibers is still low as indicated in Table 2.} The continuity of the fiber is not explicitly considered in the optimization framework. Additionally, the shortest length of printable fiber is constrained by the hardware design of the printer head, specifically the distance between the nozzle and the cutter, which limits the current fiber coverage and fiber volume fraction. \rev{}{At the same time, we need to minimize the creation of small substrate sections where fibers cannot be effectively printed.} Moreover, the quality of fiber placement is highly dependent on the substrate's printing quality. We plan to include force sensors to provide feedback to enhance the manufacturing quality of CFRTP composites.

\section*{Acknowledgments}
The project is partially supported by the chair professorship fund at the University of Manchester and the UK Engineering and Physical Sciences Research Council (EPSRC) Fellowship Grant (Ref.\#: 
EP/X032213/1). The authors would like to thank Jon Baxendale and Cristian Lira of the UK National Composites Centre for their valuable comments during this research.

\bibliographystyle{elsarticle-num} 
\bibliography{references.bib}

\end{document}